\PassOptionsToPackage{usenames, dvipsnames}{color}
\documentclass[11pt]{article}
\usepackage[letterpaper, left=1.5in, right=1.5in, top=1.5in, bottom=1.5in]{geometry}
\usepackage{pdflscape}
\usepackage{afterpage}
\usepackage[rm,bf]{titlesec}
\usepackage{algorithm}
\usepackage{algpseudocode}
\usepackage{booktabs}
\usepackage{tikz}
\usepackage{float}
\usepackage{threeparttable, booktabs}

\usepackage{enumitem}

\usepackage{amsmath}
\usepackage{amsthm}
\usepackage{multirow}
\usepackage{hyphenat}
\usepackage{graphicx}
\usepackage[export]{adjustbox}
\usepackage{amssymb}
\usepackage{amsfonts}
\usepackage{amsmath,amsthm}
\usepackage[footnotesize,bf]{caption}
\usepackage{natbib}
\usepackage{setspace}
\usepackage[dvipsnames,usenames]{color}
\usepackage{pstricks,pstricks-add}
\usepackage{pstcol}
\usepackage{url}
	
\usepackage[usenames]{color, colortbl}
\definecolor{Gray}{gray}{0.9}
\usepackage[breaklinks, colorlinks, citecolor=blue, linkcolor=blue, urlcolor=blue]{hyperref}
\usepackage{lscape}

\newtheorem{innerhypothesis}{Hypothesis}

\newtheorem{result}{Result}

\newcommand{\DRUM}{DRU}
\newcommand{\SRUM}{SRU}
\newcommand{\OSPUM}{OSPU}
\newcommand{\DFUM}{PFU}

\newcommand{\OSP}{OSP}
\newcommand{\OSS}{OSS}
\newcommand{\SOSP}{SOSP}

\newcommand{\SDB}{SDR}

\newpsobject{grilla}{psgrid}{subgriddiv=1,griddots=10,gridlabels=6pt}

\begin{document}
\bibliographystyle{elsart-harv}
\title{Testing the simplicity of strategy-proof mechanisms\thanks{This research was partially funded by NSF grant \#1458541. We are grateful to Olivier Bochet, Robert Hammond and Peter McGee for comments. This paper benefited from comments at 2021 North American Economic Science Association Meetings, SAET 2022, Granada Workshop on Collective Decisions, and seminar at the Center for Behavioral Institution Design at New York University, Abu Dhabi. We thank Ada Kovaliukaite and Valon Vitaku for conducting experiments. All errors are our own.}}
\date{\today}

\author{ Alexander L. Brown,\ \
 Daniel G. Stephenson,\  and Rodrigo A. Velez\thanks{Brown: Texas A\&M University \href{mailto:alexbrown@tamu.edu}{alexbrown@tamu.edu}; \href{http://people.tamu.edu/alexbrown}{http://people.tamu.edu/alexbrown}; Stephenson: Virginia Commonwealth University \href{mailto:stephensod@vcu.edu}{stephensod@vcu.edu};\\ \href{http://www.danielgstephenson.com}{http://www.danielgstephenson.com}\\ Velez: University of Texas at San Antonio; \href{mailto:rodrigo.velez@utsa.edu}{rodrigo.velez@utsa.edu}\\ \href{https://sites.google.com/site/rodrigoavelezswebpage/home}{https://sites.google.com/site/rodrigoavelezswebpage/home}.} }
\maketitle

\begin{abstract}
\begin{singlespace}
This paper experimentally evaluates four mechanisms intended to achieve the Uniform outcome in rationing problems \citep{Sprumont-1991}. Our benchmark is the dominant-strategy, direct-revelation mechanism of the Uniform rule. A strategically equivalent mechanism that provides non-binding feedback during the reporting period greatly improves performance. A sequential revelation mechanism produces modest improvements despite not possessing dominant strategies. A novel, obviously strategy-proof mechanism, devised by \citet{Arribillaga-et-al-2019}, does not improve performance. We characterize each alternative to the direct mechanism, finding general lessons about the advantages of real-time feedback and sequentiality of play as well as the potential shortcomings of an obviously strategy-proof mechanism.
\end{singlespace}
\medskip
\begin{singlespace}

\medskip

\medskip

\end{singlespace}
\end{abstract}


\thispagestyle{empty}
\setstretch{1.1}
\setcounter{page}{0}
\break

\section{Introduction}\label{Sec:Intro}

What ensures implementation of a social choice function? For over forty years, the gold standard in mechanism design was strategy-proofness, truthful implementation in dominant strategies. This standard persisted despite clear and well-known evidence that dominant strategies have a varied record of predicting behavior in dominant strategy games. While frequencies of dominant strategy play are low in the second-price auction \citep{Kegel-et-al-1987-Eca}, the Top Trading Cycles, and the Deferred Acceptance mechanisms \citep{CHEN-Sonmez-2006-JET}, agents tend to drop close to their values in an English auction \citep{Kegel-et-al-1987-Eca}. Based on these observations, \citet{Li-AER-17} identified Obvious Strategy-Proofness (OSP), a mechanism property that differentiates the second-price and English auctions, two mechanisms that are strategically equivalent. While the idea of more robust implementation of social choice functions has always had interest in mechanism design \citep[e.g.,][]{Saijo-et-al-2007-TE, Bergemann-Morris-2011-GEB}, following \citeauthor{Li-AER-17}'s seminal contribution, there has been great interest in designing OSP mechanisms \citep[e.g.,][]{Arribillaga-et-al-2019}, in determining the extent to which OSP is responsible for the difference between the second-price and English auctions \citep[e.g.][]{MCGEE2019355,Breit-Sch-2022-ExpE}, and in identifying other properties that also differentiate strategy-proof mechanisms for the implementation of a social choice function \citep{Pycia-Troyan-2019}.

This paper provides an experimental study that compares four mechanisms intended to obtain the \emph{same} outcomes in a rationing problem with satiable preferences \citep{Sprumont-1991}. The mechanisms we test are inspired by the literatures on dominant-strategy implementation \citep{Sprumont-1991}, obviously-dominant strategy implementation \citep{Li-AER-17,Arribillaga-et-al-2019}, perfect Bayesian robust implementation \citep{Schummer-Velez-2019}, and feedback-augmented/cooperative implementation \citep{BOCHET-Tumme-JET-2020,stephenson2022assignment}.

We make hypothesis about the performance and behavior in the mechanisms we test based on standard theory and in the formalism introduced by \citet{Li-AER-17} as refined by \citet{Pycia-Troyan-2019}.

 

\citet{Li-AER-17}'s work focuses on the removal of contingent reasoning about other agents' strategies. \cite{Pycia-Troyan-2019} look to additionally characterize the simplicity of strategies based on how far ahead down an agent's intended path of play they must foresee in order to realize their plan is dominant. Their basic definition is a $k$-step Simply Dominant intended path of play at a node: at most $k$ consecutive actions that can be identified as dominant without the need of contingent reasoning (see See Sec.~\ref{Sec:Eq-predictions} and \cite{Pycia-Troyan-2019} for details). Intuitively, a $k$-step Simply Dominant plan is simpler than a $(k+1)$-step Simply Dominant plan. They both require no contingent reasoning, but the former requires the agent trusts their capacity to carry out a shorter plan. \citet{Li-AER-17}'s \OSP\ corresponds to requiring intended paths of play be $\infty$-step Simply Dominant.

It is quite rare for any non-dictatorial mechanism to satisfy \citet{Pycia-Troyan-2019}'s simplicity properties at every node.\footnote{Dictatorial mechanisms are simple beyond \cite{Pycia-Troyan-2019}'s simplicity categorization. Thus, they are not an ideal test of the theory. See Sec.~\ref{Sec:Eq-predictions} for a discussion.} Rationing problems with satiable preferences stand out as the sole environment for which a prominent 
non-dictatorial social choice function 
can be implemented by a mechanism satisfying %
some of the most strict requirements in that categorization.  Indeed, there is a Uniform mechanism that admits 1-step Simply Dominant strategic plans \citep{Arribillaga-et-al-2019}. Some of the actions called for by optimal plans in this mechanism are  0-step Simply Dominant, the most stringent requirement. We refer to this mechanism as \OSPUM. To our knowledge, we are the first to test any such mechanism experimentally.

Besides \OSPUM, we look at alternatives to implement the Uniform outcome inspired by other literature.  A sequential mechanism (\SRUM) allows one agent to go first and make report, leaving the second agent to complete their report under complete certainty. While it is not strategy-proof due to the lack of a dominant strategy for the first player, its only perfect Bayesian equilibrium outcome is Uniform \citep{Schummer-Velez-2019}. Further, experimental studies suggest the removal of uncertainty---all else being equal---is associated with a higher level of rationality \citep{martinez2019failures}, so we would expect better decision-making for the second mover, with ambiguous consequences for the first. 

We also test a direct Uniform mechanism that provides feedback during the reporting period (\DFUM). This mechanism allows subjects to observe the tentative report currently selected by their counterpart and all possible outcomes under this report. Assignments are exclusively determined by the finalized reports at the end of the reporting period, so tentative reports are costless and non-binding. Conventional implementations of direct mechanisms often already have a reporting period during which participants can freely select and adjust their reports. The new feature here is the provision of feedback during the reporting period. Such feedback is shown to improve equilibration in school choice mechanisms in laboratory experiments \citep{stephenson2022assignment}.\footnote{Real-time feedback may be logistically feasible in some internet-based deployment of mechanisms. For instance, similar feedback has been provided in actual school-choice environments in Wake County, NC \citep{Dur-et-al-2018-AEJEP} and Inner Mongolia, China \citep{kang2023dynamic}.
} Also, if agents are able to use this type of limited communication to coordinate, they may have aligned incentives to coordinate on the Uniform outcome \citep{BOCHET-Tumme-JET-2020,Bochet-Tummenassan-2020}.

Our main results compare \OSPUM, \SRUM, and \DFUM\ with the baseline \DRUM . The feedback mechanism, \DFUM, greatly improves performance. The sequentiality of \SRUM\ modestly improves performance. The additional simplicity of \OSPUM\ does not improve performance. To understand the reasons behind this ranking, we characterize behavior in the different mechanisms and identify possible key driving forces.

Subjects in \DRUM\ achieve the Uniform outcome in about half of all instances, though this outcome is often achieved via non-dominant strategies. Profiles of observed play in \DRUM\ often resemble weakly-dominated Nash equilibria, so subjects seem to be fairly successful in making correct conjectures about the play of others and responding accordingly. Despite differing incentives, first-mover subjects in \SRUM\ play quite similarly to subjects in \DRUM. By contrast, second movers seem to understand their informational advantage and respond accordingly. Consequently, \SRUM\ has no worse, and sometimes better, performance than \DRUM. 

The tentative reports in \DFUM\ are highly correlated with final reports. This gives both subjects in \DFUM\ a very similar informational advantage to the second mover in \SRUM. Most agents end up maximizing their payoffs in \DFUM. Additionally, dominant strategies seem to have a greater appeal because tentative reports vary and remain uncertain during the reporting period in \DFUM. As a result, the frequency of dominant strategies is higher for \DFUM\ than for second-movers in \SRUM. Since a single agent choosing a dominant strategy leads to the Uniform outcome when the other agent best responds, \DFUM\ achieves the highest rates of Uniform outcomes.

Contrary to the expectations of  \citeauthor{Li-AER-17} 
and \cite{Pycia-Troyan-2019}'s theory, \OSPUM\ does not lead to better outcomes than \DRUM. At the node level, rates of actions consistent with dominant strategies are higher in \OSPUM\ than any other mechanism. This advantage at the node level is reduced in terms of the frequency of realized equilibrium paths as the \OSPUM\ is the only mechanism to contain multiple node per player. The mechanism is also unique in that it often prevents the existence of Nash equilibria where the Uniform outcome is achieved through weakly-dominated reports. These two main issues lower \OSPUM\ performance relative to \DRUM. 

The remainder of the paper is organized as follows. Sec.~\ref{Section-UR} introduces rationing problems and the mechanisms we test. Sec.~\ref{Sec-ED-P} presents our experimental design. Sec.~\ref{Sec:pred-hyp} develops our hypotheses. Sec.~\ref{Sec.Results} presents our results. Sec.~\ref{Sec-Discussion} concludes.
\section{Uniform Rationing and Mechanisms}\label{Section-UR}

\subsection{Uniform rationing}

We consider rationing problems with satiable preferences \citep{Benassy-1982,Sprumont-1991}.  For concreteness and anticipating our experimental design we assume there are two agents $N=\{1,2\}$ and twenty units of a commodity to distribute. Agent~$i$ has a preferred amount of the good $\theta_i\in\{0,1,...,20\}=\Theta_i$. We refer to this amount as the agent's peak. The agent loses utility as their assignment moves away from this ideal consumption. 

An example is an instructor who needs to assign twenty hours of grading and twenty hours of tutoring among two teaching assistants. Each teaching assistant needs to be assigned a total of twenty hours of work.  Thus, the allocation can be reduced to decide the number of tutoring hours each TA serves. If teaching assistants have convex preferences on bundles of grading and tutoring, their preferences on tutoring are satiable.

We assume that if agent~$i$ is assigned an amount $x\in\{1,...,20\}$, her payoff is
\begin{equation}u_i(x,\theta_i)=K-|\theta_i-x|,\label{EQ:utility}\end{equation}
and that agents are expected payoff maximizers. We choose $K=20$ to avoid the possibility of bankruptcy in our experiment.

\subsection{Uniform rule}

A social choice function selects an allocation for each state. The most prominent scf in our environment is the Uniform Rule \citep{Benassy-1982,Sprumont-1991}, which determines the allocation for a given state $\theta=(\theta_1,\theta_2)\in \Theta=\Theta_1\times\Theta_2$. One can describe this rule as follows. It determines if there is too much or too little demand of the good and assigns responsibility for this to the agent(s) whose demand is below or above the average assignment, in this case $10$. If both agents are responsible for the mistmatch of demand and supply, both get $10$. Otherwise, the agent who is not responsible for the mismatch receives their demand, and the other agent receives the residual claim.\footnote{Formally, if $\theta_1+\theta_2=20$, the agents receive their preferred amounts; if $\theta_1+\theta_2>20$, there is a unique $\lambda\geq0$ for which $\min\{\theta_1,\lambda\}+\min\{\theta_2,\lambda\}=20$ and agent~$i$ receives $\min\{\theta_i,\lambda\}$; if $\theta_1+\theta_2<20$, there is a unique $\lambda\geq0$ for which $\max\{\theta_1,\lambda\}+\max\{\theta_2,\lambda\}=20$ and agent~$i$ receives $\max\{\theta_i,\lambda\}$.} We denote this recommendation by $U(\theta)$. 
\begin{figure}[t]
\centering
\begin{pspicture}(0,0.8)(5.2,6)
\rput[c](3,1.2){\footnotesize Report/Award Subject 1}
\rput[c]{90}(0.2,4){\footnotesize Report/Award Subject 2}
\rput(1,1.7){$\mbox{\footnotesize$0$}$}
\rput(2,1.7){$\mbox{\footnotesize$5$}$}
\rput(3,1.7){$\mbox{\footnotesize$10$}$}
\rput(4,1.7){$\mbox{\footnotesize$15$}$}
\rput(5,1.7){$\mbox{\footnotesize$20$}$}
\rput[r](0.8,2){$\mbox{\footnotesize$0$}$}
\rput[r](0.8,3){$\mbox{\footnotesize$5$}$}
\rput[r](0.8,4){$\mbox{\footnotesize$10$}$}
\rput[r](0.8,5){$\mbox{\footnotesize$15$}$}
\rput[r](0.8,6){$\mbox{\footnotesize$20$}$}
\psline[linecolor=red](1,6)(5,2)
\psline[linewidth=4pt]{->}(4,2)(4,2.9)
\psline[linewidth=4pt]{->}(2,6)(2,5.1)
\psline[linewidth=4pt]{->}(1,5)(1.9,5)
\psline[linewidth=4pt]{->}(5,3)(4.1,3)
\psline[linewidth=4pt]{->}(3,2)(3,3.9)
\psline[linewidth=4pt]{->}(1,4)(2.9,4)
\psline[linewidth=4pt]{->}(1,2)(2.9,3.9)
\psline[linewidth=4pt]{->}(3,6)(3,4.1)
\psline[linewidth=4pt]{->}(5,4)(3.1,4)
\psline[linewidth=4pt]{->}(5,6)(3.1,4.1)
\psdots[dotstyle=square](1,2)
(1.2,2)
(1.4,2)
(1.6,2)
(1.8,2)
(2,2)
(2.2,2)
(2.4,2)
(2.6,2)
(2.8,2)
(3,2)
(3.2,2)
(3.4,2)
(3.6,2)
(3.8,2)
(4,2)
(4.2,2)
(4.4,2)
(4.6,2)
(4.8,2)
(5,2)
(1,2.2)
(1.2,2.2)
(1.4,2.2)
(1.6,2.2)
(1.8,2.2)
(2,2.2)
(2.2,2.2)
(2.4,2.2)
(2.6,2.2)
(2.8,2.2)
(3,2.2)
(3.2,2.2)
(3.4,2.2)
(3.6,2.2)
(3.8,2.2)
(4,2.2)
(4.2,2.2)
(4.4,2.2)
(4.6,2.2)
(4.8,2.2)
(5,2.2)
(1,2.4)
(1.2,2.4)
(1.4,2.4)
(1.6,2.4)
(1.8,2.4)
(2,2.4)
(2.2,2.4)
(2.4,2.4)
(2.6,2.4)
(2.8,2.4)
(3,2.4)
(3.2,2.4)
(3.4,2.4)
(3.6,2.4)
(3.8,2.4)
(4,2.4)
(4.2,2.4)
(4.4,2.4)
(4.6,2.4)
(4.8,2.4)
(5,2.4)
(1,2.6)
(1.2,2.6)
(1.4,2.6)
(1.6,2.6)
(1.8,2.6)
(2,2.6)
(2.2,2.6)
(2.4,2.6)
(2.6,2.6)
(2.8,2.6)
(3,2.6)
(3.2,2.6)
(3.4,2.6)
(3.6,2.6)
(3.8,2.6)
(4,2.6)
(4.2,2.6)
(4.4,2.6)
(4.6,2.6)
(4.8,2.6)
(5,2.6)
(1,2.8)
(1.2,2.8)
(1.4,2.8)
(1.6,2.8)
(1.8,2.8)
(2,2.8)
(2.2,2.8)
(2.4,2.8)
(2.6,2.8)
(2.8,2.8)
(3,2.8)
(3.2,2.8)
(3.4,2.8)
(3.6,2.8)
(3.8,2.8)
(4,2.8)
(4.2,2.8)
(4.4,2.8)
(4.6,2.8)
(4.8,2.8)
(5,2.8)
(1,3)
(1.2,3)
(1.4,3)
(1.6,3)
(1.8,3)
(2,3)
(2.2,3)
(2.4,3)
(2.6,3)
(2.8,3)
(3,3)
(3.2,3)
(3.4,3)
(3.6,3)
(3.8,3)
(4,3)
(4.2,3)
(4.4,3)
(4.6,3)
(4.8,3)
(5,3)
(1,3.2)
(1.2,3.2)
(1.4,3.2)
(1.6,3.2)
(1.8,3.2)
(2,3.2)
(2.2,3.2)
(2.4,3.2)
(2.6,3.2)
(2.8,3.2)
(3,3.2)
(3.2,3.2)
(3.4,3.2)
(3.6,3.2)
(3.8,3.2)
(4,3.2)
(4.2,3.2)
(4.4,3.2)
(4.6,3.2)
(4.8,3.2)
(5,3.2)
(1,3.4)
(1.2,3.4)
(1.4,3.4)
(1.6,3.4)
(1.8,3.4)
(2,3.4)
(2.2,3.4)
(2.4,3.4)
(2.6,3.4)
(2.8,3.4)
(3,3.4)
(3.2,3.4)
(3.4,3.4)
(3.6,3.4)
(3.8,3.4)
(4,3.4)
(4.2,3.4)
(4.4,3.4)
(4.6,3.4)
(4.8,3.4)
(5,3.4)
(1,3.6)
(1.2,3.6)
(1.4,3.6)
(1.6,3.6)
(1.8,3.6)
(2,3.6)
(2.2,3.6)
(2.4,3.6)
(2.6,3.6)
(2.8,3.6)
(3,3.6)
(3.2,3.6)
(3.4,3.6)
(3.6,3.6)
(3.8,3.6)
(4,3.6)
(4.2,3.6)
(4.4,3.6)
(4.6,3.6)
(4.8,3.6)
(5,3.6)
(1,3.8)
(1.2,3.8)
(1.4,3.8)
(1.6,3.8)
(1.8,3.8)
(2,3.8)
(2.2,3.8)
(2.4,3.8)
(2.6,3.8)
(2.8,3.8)
(3,3.8)
(3.2,3.8)
(3.4,3.8)
(3.6,3.8)
(3.8,3.8)
(4,3.8)
(4.2,3.8)
(4.4,3.8)
(4.6,3.8)
(4.8,3.8)
(5,3.8)
(1,4)
(1.2,4)
(1.4,4)
(1.6,4)
(1.8,4)
(2,4)
(2.2,4)
(2.4,4)
(2.6,4)
(2.8,4)
(3,4)
(3.2,4)
(3.4,4)
(3.6,4)
(3.8,4)
(4,4)
(4.2,4)
(4.4,4)
(4.6,4)
(4.8,4)
(5,4)
(1,4.2)
(1.2,4.2)
(1.4,4.2)
(1.6,4.2)
(1.8,4.2)
(2,4.2)
(2.2,4.2)
(2.4,4.2)
(2.6,4.2)
(2.8,4.2)
(3,4.2)
(3.2,4.2)
(3.4,4.2)
(3.6,4.2)
(3.8,4.2)
(4,4.2)
(4.2,4.2)
(4.4,4.2)
(4.6,4.2)
(4.8,4.2)
(5,4.2)
(1,4.4)
(1.2,4.4)
(1.4,4.4)
(1.6,4.4)
(1.8,4.4)
(2,4.4)
(2.2,4.4)
(2.4,4.4)
(2.6,4.4)
(2.8,4.4)
(3,4.4)
(3.2,4.4)
(3.4,4.4)
(3.6,4.4)
(3.8,4.4)
(4,4.4)
(4.2,4.4)
(4.4,4.4)
(4.6,4.4)
(4.8,4.4)
(5,4.4)
(1,4.6)
(1.2,4.6)
(1.4,4.6)
(1.6,4.6)
(1.8,4.6)
(2,4.6)
(2.2,4.6)
(2.4,4.6)
(2.6,4.6)
(2.8,4.6)
(3,4.6)
(3.2,4.6)
(3.4,4.6)
(3.6,4.6)
(3.8,4.6)
(4,4.6)
(4.2,4.6)
(4.4,4.6)
(4.6,4.6)
(4.8,4.6)
(5,4.6)
(1,4.8)
(1.2,4.8)
(1.4,4.8)
(1.6,4.8)
(1.8,4.8)
(2,4.8)
(2.2,4.8)
(2.4,4.8)
(2.6,4.8)
(2.8,4.8)
(3,4.8)
(3.2,4.8)
(3.4,4.8)
(3.6,4.8)
(3.8,4.8)
(4,4.8)
(4.2,4.8)
(4.4,4.8)
(4.6,4.8)
(4.8,4.8)
(5,4.8)
(1,5)
(1.2,5)
(1.4,5)
(1.6,5)
(1.8,5)
(2,5)
(2.2,5)
(2.4,5)
(2.6,5)
(2.8,5)
(3,5)
(3.2,5)
(3.4,5)
(3.6,5)
(3.8,5)
(4,5)
(4.2,5)
(4.4,5)
(4.6,5)
(4.8,5)
(5,5)
(1,5.2)
(1.2,5.2)
(1.4,5.2)
(1.6,5.2)
(1.8,5.2)
(2,5.2)
(2.2,5.2)
(2.4,5.2)
(2.6,5.2)
(2.8,5.2)
(3,5.2)
(3.2,5.2)
(3.4,5.2)
(3.6,5.2)
(3.8,5.2)
(4,5.2)
(4.2,5.2)
(4.4,5.2)
(4.6,5.2)
(4.8,5.2)
(5,5.2)
(1,5.4)
(1.2,5.4)
(1.4,5.4)
(1.6,5.4)
(1.8,5.4)
(2,5.4)
(2.2,5.4)
(2.4,5.4)
(2.6,5.4)
(2.8,5.4)
(3,5.4)
(3.2,5.4)
(3.4,5.4)
(3.6,5.4)
(3.8,5.4)
(4,5.4)
(4.2,5.4)
(4.4,5.4)
(4.6,5.4)
(4.8,5.4)
(5,5.4)
(1,5.6)
(1.2,5.6)
(1.4,5.6)
(1.6,5.6)
(1.8,5.6)
(2,5.6)
(2.2,5.6)
(2.4,5.6)
(2.6,5.6)
(2.8,5.6)
(3,5.6)
(3.2,5.6)
(3.4,5.6)
(3.6,5.6)
(3.8,5.6)
(4,5.6)
(4.2,5.6)
(4.4,5.6)
(4.6,5.6)
(4.8,5.6)
(5,5.6)
(1,5.8)
(1.2,5.8)
(1.4,5.8)
(1.6,5.8)
(1.8,5.8)
(2,5.8)
(2.2,5.8)
(2.4,5.8)
(2.6,5.8)
(2.8,5.8)
(3,5.8)
(3.2,5.8)
(3.4,5.8)
(3.6,5.8)
(3.8,5.8)
(4,5.8)
(4.2,5.8)
(4.4,5.8)
(4.6,5.8)
(4.8,5.8)
(5,5.8)
(1,6)
(1.2,6)
(1.4,6)
(1.6,6)
(1.8,6)
(2,6)
(2.2,6)
(2.4,6)
(2.6,6)
(2.8,6)
(3,6)
(3.2,6)
(3.4,6)
(3.6,6)
(3.8,6)
(4,6)
(4.2,6)
(4.4,6)
(4.6,6)
(4.8,6)
(5,6)
\end{pspicture}
\caption{Uniform rule.}
\label{Fig:Uniform1}
\end{figure}
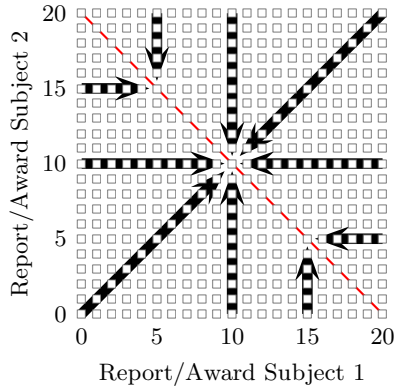

Figure~\ref{Fig:Uniform1} shows a useful graphical representation of  $U$ due to \citet{Moulin-2017-TE}. The type space, $\Theta_1\times\Theta_2$, is the cross product $\{0,...,20\}\times \{0,...,20\}$. One can see a rule as a map from the type space into the set of feasible allotments, i.e., the diagonal with slope~$-1$.  Each $(\theta_1,\theta_2)$ in which both peaks are at most $10$, or both peaks are at least $10$, are  mapped to $(10,10)$. For each $(\theta_1,\theta_2)$ in which one peak is below $10$ and one is above $10$, one can find $U(\theta_1,\theta_2)$ by horizontally or vertically ``projecting'' the profile on the diagonal of feasible awards. The direction of the projection is the one that produces the award closest to equal division. For instance, consider the profile of peaks $(17,5)$. This profile belongs to the lower triangle that forms the right lower quadrant of the type space. The Uniform outcome for this profile is obtained by projecting vertically on the feasible diagonal. It results in award $(15,5)$. By contrast, the recommendation for profile $(3,13)$ is found by projecting horizontally on the feasible diagonal. In this case the Uniform outcome is $(7,13)$.

The Uniform rule has multiple desirable properties. It is envy-free, i.e., no agent prefers the allotment of another agent to her own \citep[c.f.,][]{Sprumont-1991}. It is Pareto efficient, i.e., no agent can be better off without any other agent being worse off \citep[c.f.,][]{Sprumont-1991}.\footnote{In this environment, an allocation is Pareto efficient whenever all agents consume on the same side of their peak.} It is non-bossy, i.e., no agent can change the outcome without changing her own welfare \cite[c.f.,][]{Schummer-Velez-2019}. Finally, it is strategy-proof, i.e., for each $\theta\in\Theta$, each $i\in N$, and each $\theta_i'\in\Theta_i$, $u_i(U(\theta)|\theta_i)\geq u_i(U(\theta_i',\theta_{-i})|\Theta_i)$ \citep[c.f.,][]{Sprumont-1991}.

\subsection{Mechanisms}\label{Sec-mechanisms}

We consider four different mechanisms that, according to standard theoretical solutions, should obtain the Uniform outcome. 
\begin{itemize}
\item \textbf{Direct Revelation Uniform Rationing (\DRUM)}: Simultaneously ask agents for their types. Then assign the Uniform allocation for the reports. 
\item \textbf{Sequential Revelation Uniform Rationing (\SRUM)}: Sequentially ask agents for their types and reveal the partial reports to agents down the line (as in a round table or roll call). Then assign the Uniform allocation for the reports. 
    
\item \textbf{Obviously Strategy-proof Uniform Rationing (\OSPUM )}: The assignment is done by means of a clock-auction type procedure. To start, let the initial temporary assignment be $x_1^0=x_2^0=10$.

      \begin{enumerate}
        \item \textbf{Step 0}: Inform agents of their temporary assignment. Ask each $i\in N$ to choose between $x_i^0-1$, $x_i^0$, and $x_i^0+1$. Let $x_1^1$ and $x_2^1$ be their choices. If some agent chooses her initial temporary assignment or $x_1^1+x_2^1\neq 20$, they get $x^0_1$ and $x^0_1$, respectively.  Otherwise, revise the temporary assignment to $x_1^1$ and $x_2^1$.  Let $\Delta_i=x_i^1-x_i^0$. Note that $\Delta_1$ and $\Delta_2$ are both different from zero and $\Delta_1+\Delta_2=0$ .

        \item \textbf{Step t=1,2,...}: At the beginning of this step, temporary allotments are $x^t_1>0$ and $x^t_2>0$. Inform agents that they have been temporarily assigned $x_i^1$. Inform agent~$i$ that unless one of the agents opts out, their temporary assignment will be revised to $x^t_i+\Delta_i$. Provide a fixed time to make this decision. If at least one agent opts out,  $x^t_1$ and $x^t_2$ becomes the final assignment and the game finishes. If no agent opts out, the temporary allotments are revised to $x^{t+1}_i=x^t_i+\Delta_i$. The game finishes if one temporary assignment becomes zero. At this point the temporary assignment becomes final. Otherwise, the game continues to Step $t+1$.
      \end{enumerate}
\item \textbf{Pre-play Feedback Uniform Rationing (\DFUM)}: Ask agents directly for their types. Agents can adjust their reports over a finite period spanning the time interval $\left[0,T\right]\subseteq\mathbb{R}_{+}$. During the reporting period, agents are informed about their tentative assignments under the currently selected reports. Finalized assignments are determined by the finalized reports selected at the end of the reporting period. (\DFUM\ is identical to \DRUM, except for the feedback provided during the reporting period.)
\end{itemize}

\section{Experimental Design and Procedures}\label{Sec-ED-P}

\subsection{Design}\label{Sec-Exp-design}
Our experimental design has four treatments that implement each one of the four mechanisms described in Sec.~\ref{Sec-mechanisms}: \DRUM , \SRUM , \OSPUM , and \DFUM . The treatments were administered between subjects: each experimental session implemented only one of the four mechanisms.

An experimental session consisted of 12 periods. At the beginning of each period, subjects in each pair were assigned types as shown in Table~\ref{tab:TypeAssignments}. Subjects observe both their own type and their partner's type. During each period, subjects participated in the assignment mechanism. At the end of each period, payoffs were determined by a subject's type and their assignment following the guidelines of their respective mechanism. 
Subjects were randomly rematched after each period.\footnote{The \SRUM\ was the only treatment to feature asymmetric roles for each subject due to the underlying design of the mechanism. To increase familiarity with the mechanism, subjects preserved their role over all periods, meaning an individual subject was either a first or second mover for the entirety of the experiment. Under this scheme, a subject could only be randomly matched with half the subjects in any experiment session. For the sake of comparability, we follow this matching scheme across all treatments, randomly dividing subjects into two arbitrary groups and matching accordingly for the other symmetric mechanisms.}

\begin{table}
\begin{centering}
\begin{tabular}{ccccccccccccc}
\hline
Period & 1 & 2 & 3 & 4 & 5 & 6 & 7 & 8 & 9 & 10 & 11 & 12\\
Valuation Pair & 1 & 2 & 3 & 4 & 5 & 6 & 1 & 2 & 3 & 4 & 5 & 6 \\
\hline
\noalign{\vskip0.1cm}
Subject A's Type & 3 & 15 & 16 & 3 & 5 & 9 & 4 & 16 & 4 & 13 & 17 & 11\\
\noalign{\vskip0.05cm}
Subject B's Type & 4 & 16 & 4 & 13 & 17 & 11 & 3 & 15 & 16 & 3 & 5 & 9\\
\noalign{\vskip0.1cm}
Allocation A & 10 & 10 & 16 & 7 & 5 & 9 & 10 & 10 & 4 & 13 & 15 & 11\\
\noalign{\vskip0.05cm}
Allocation B & 10 & 10 & 4 & 13 & 15 & 11 & 10 & 10 & 16 & 7 & 5 & 9\\
\noalign{\vskip0.1cm}
Payoff A & 13 & 15 & 20 & 16 & 20 & 20 & 14 & 14 & 20 & 20 & 18 & 20\\
\noalign{\vskip0.05cm}
Payoff B & 14 & 14 & 20 & 20 & 18 & 20 & 13 & 15 & 20 & 16 & 20 & 20\\
\hline
\end{tabular}
\par\end{centering}
\caption{Type assignments, allocations, and payoffs in Uniform outcome by period.\label{tab:TypeAssignments}}
\end{table}

\begin{figure}[t]
\begin{minipage}[t]{0.5\textwidth}
\begin{centering}
\begin{tabular}{c}
\tabularnewline
\includegraphics[width=\textwidth]{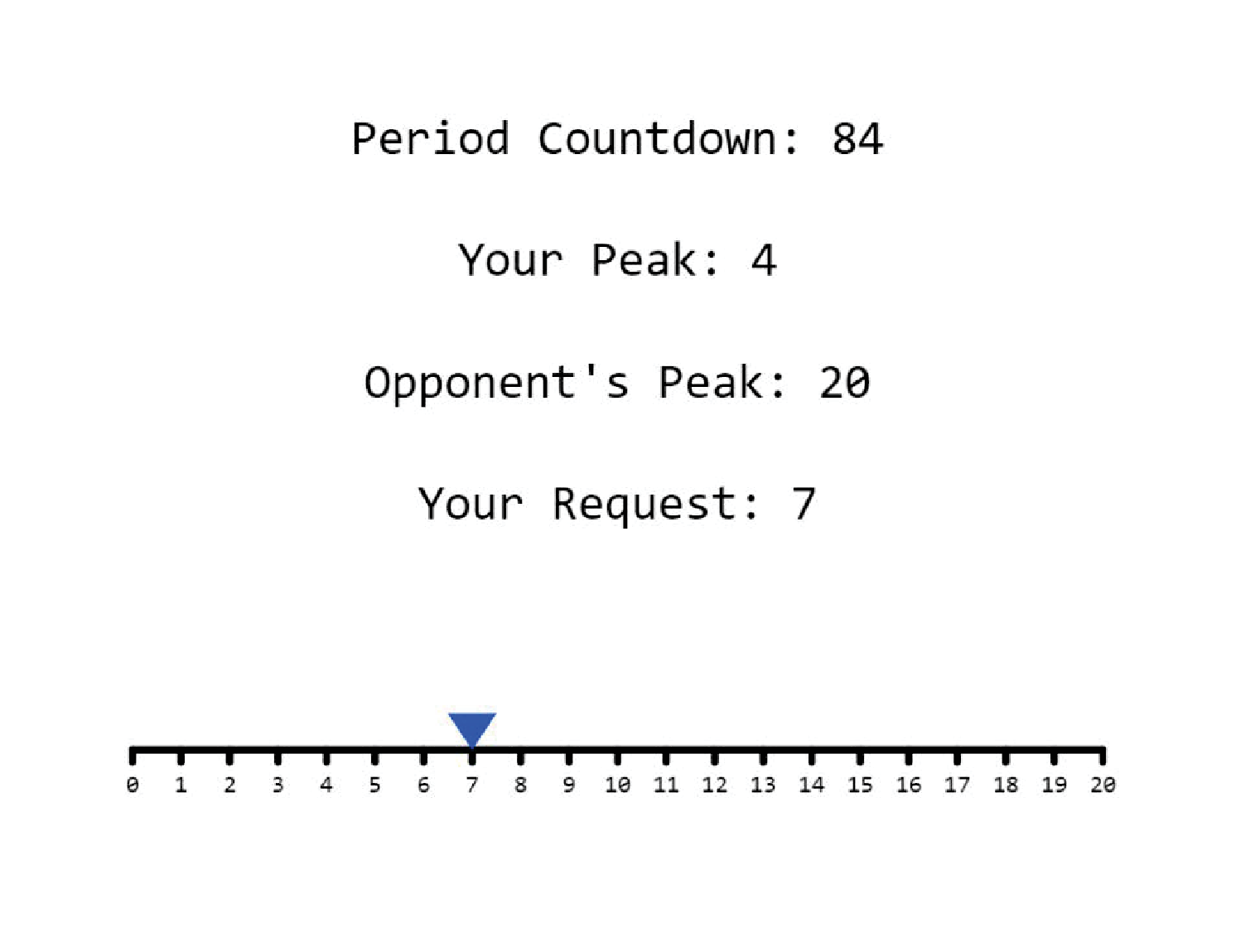}\tabularnewline
\tabularnewline
\end{tabular}
\par\end{centering}
\caption{Direct Revelation Uniform. (\DRUM)\label{fig:ScreenshotDRU}}
\end{minipage}
\begin{minipage}[t]{0.5\textwidth}
\begin{centering}
\begin{tabular}{c}
\tabularnewline
\includegraphics[width=\textwidth]{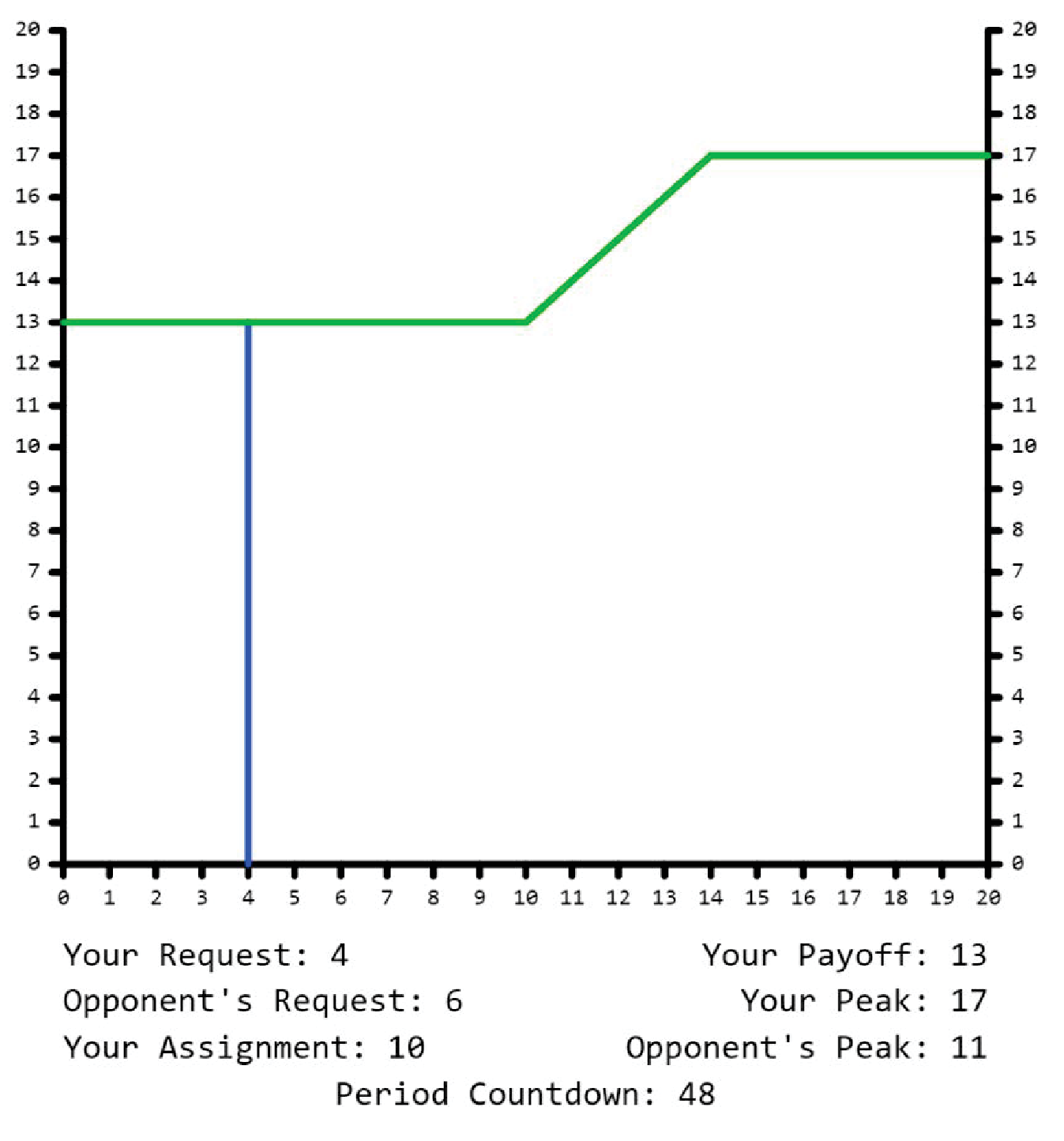}\tabularnewline
\tabularnewline
\end{tabular}
\par\end{centering}
\caption{Pre-play Feedback Uniform Rationing. (\DFUM)\label{fig:ScreenshotCFU}}
\end{minipage}
\end{figure}
\begin{figure}[]
\begin{centering}
\begin{tabular}{c}
\tabularnewline
\includegraphics[height=7cm, frame]{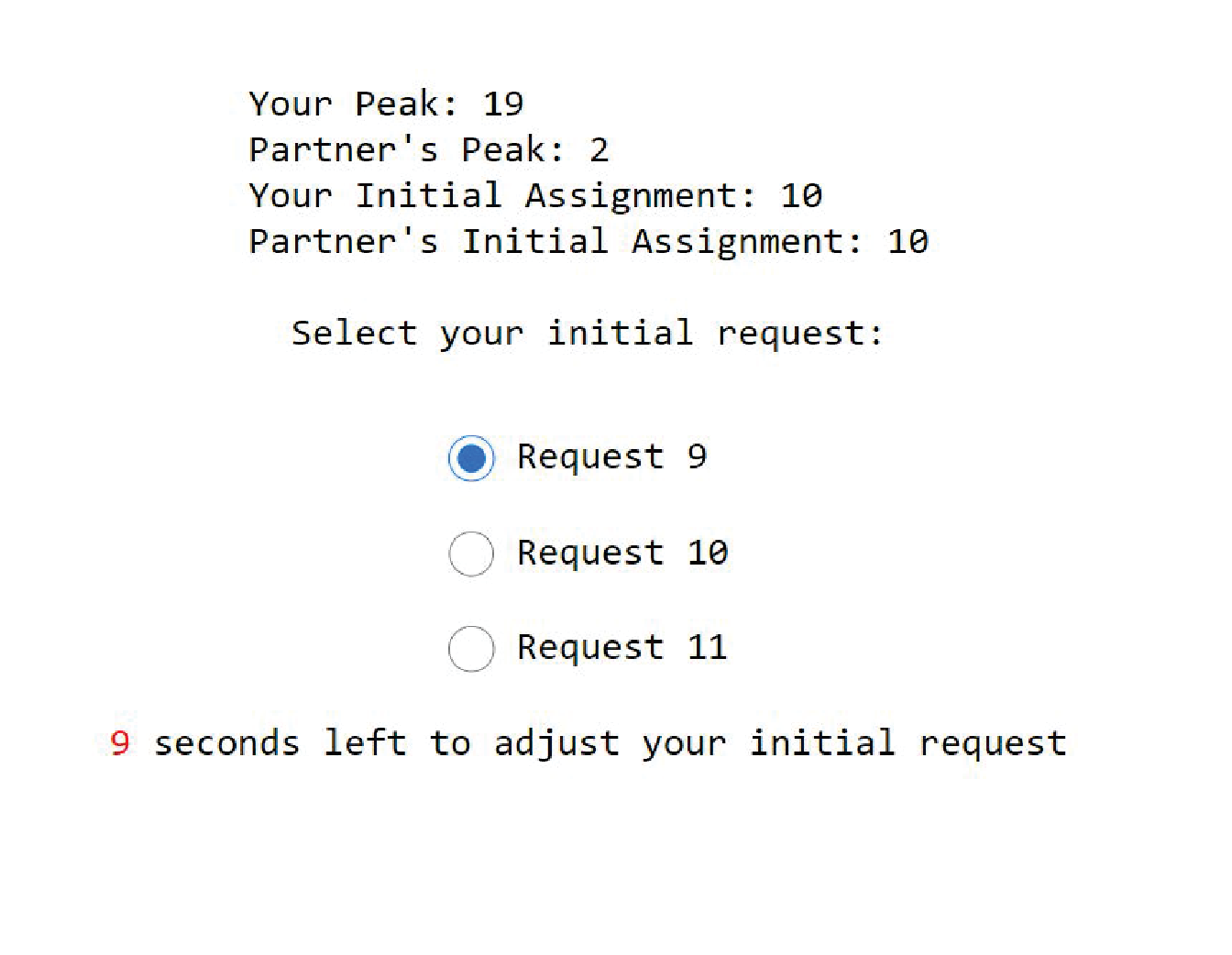}\tabularnewline
\includegraphics[height=7cm, frame]{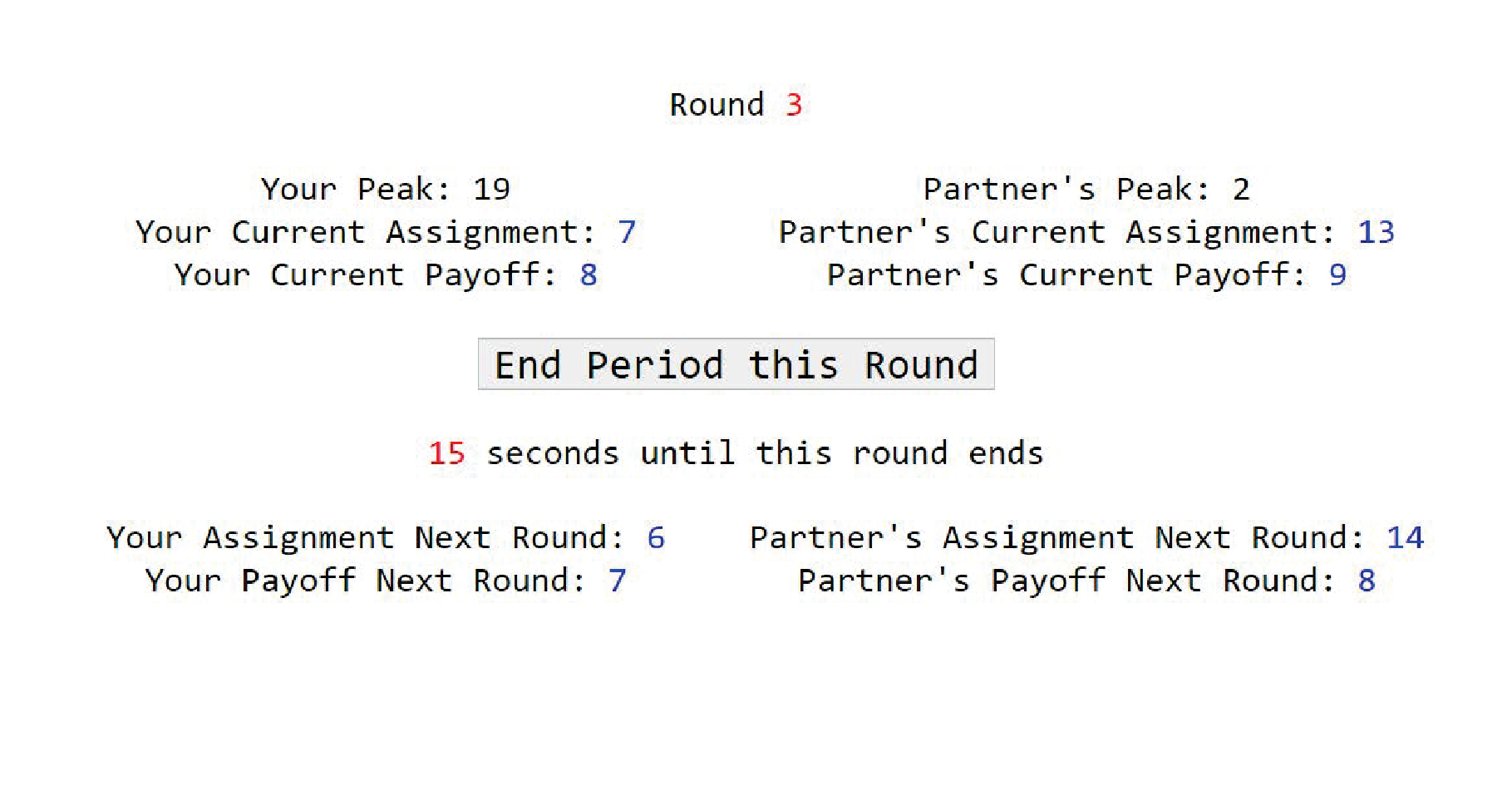}\tabularnewline
\tabularnewline
\end{tabular}
\par\end{centering}
\caption{Obviously Strategy-proof Uniform Rationing. (\OSPUM)\label{fig:ScreenshotOSP}}
\end{figure}

Figures~\ref{fig:ScreenshotDRU} and~\ref{fig:ScreenshotCFU} depict the experimental interface for \DRUM\ and \DFUM\ treatments, respectively. In \SRUM, the first mover uses an interface similar to that of \DRUM\ and the second mover uses an interface similar to that of \DFUM. Figure~\ref{fig:ScreenshotOSP} depicts the experimental interface for \OSPUM\ treatment. The first row depicts the interface for selecting initial temporary assignments. The second row depicts the interface for opting out.

A particular feature of our design is that we recorded subjects' selection (cursor position) ten times per second in the window of time in which the decision was made for \DRUM, \SRUM\, and \DFUM. A subject's current tentative report was determined by their current mouse position throughout the reporting period. A subject's finalized report was determined by their final mouse position at the end of the reporting period. Finalized assignments were exclusively based on finalized reports and earnings were exclusively based on finalized assignments.

\subsection{Procedures}\label{Sec-Exp-Procedures}
%

One group of 14 and one group of 10 subjects participated in the continuous feedback Uniform mechanism (\DFUM) sessions. Two groups of 14 and one group of 10 subjects participated in the sequential revelation Uniform mechanism (\SRUM) sessions. Three groups of 14 and one group of 10 subjects participated in the direct revelation mechanism (\DRUM) sessions. All transpired during October 2017. Five sessions of the obviously strategy-proof Uniform mechanism (\OSPUM) were conducted in June 2021. These sessions featured three groups of 10 subjects and two groups of 8 subjects, respectively.

All sessions were held at the Economic Research Laboratory (ERL) in the Economics Department at Texas A\&M University. The 160 subjects were recruited using ORSEE software \citep{greiner} from a variety of majors.  Experiments lasted about one hour. Subjects were paid based on their average earnings over all periods, using a conversion rate of $\$1=1$ ECU plus a \$10 participation payment. The average subject earnings were \$26.19.

\section{Predictions and hypotheses}\label{Sec:pred-hyp}

\subsection{Equilibrium predictions}\label{Sec:Eq-predictions}
%
Let us consider an arbitrary common prior on $\Theta$, the payoff-type space, and compare the standard game theoretical predictions for it. 
Each mechanism with the prior defines a standard game of perfect information: Simultaneous-move games for \DRUM\ and \DFUM\ (its payoff relevant move); extensive-form games for \SRUM\ and \OSPUM. Table~\ref{tab:EqPredictions} summarizes standard predictions for all these games.

The first relevant benchmark is Nash equilibrium (perfect equilibrium for extensive\hyp{}form games). All mechanisms have an equilibrium that obtains the Uniform outcome: unconditionally report true peaks in \DRUM, \SRUM\, and \DFUM; move towards peak at each node in \OSPUM. However, \SRUM\ is the only mechanism that guarantees each equilibrium outcome is Uniform \citep{Schummer-Velez-2019}. \DRUM, \OSPUM, and \DFUM\ all possess ex-post equilibria that do not obtain the Uniform outcome.\footnote{Unconditionally requesting $10$ is a Nash equilibrium in all these mechanisms independently of information structure. Under complete information, the Nash equilibria of \DRUM\ form a lattice with respect to the Pareto order with extremes equal division and the Uniform outcome \citep{BOCHET-Tumme-JET-2020}.} 

\begin{table}[]
\centering
\resizebox{1\textwidth}{!} {
\begin{threeparttable}\footnotesize
\begin{tabular}{lcccccc}
\toprule
 & (Perfect) & Dominant& Obviously &1-step Simply&0-step Simply& \\
 & Nash & strategy& Dominant &Dominant&Dominant&deterministic\\
mech. & equil. & equil. &str. plan &str. plan&str. plan&choice \\
\midrule
\DRUM & $\textbf{--}$              &$\textbf{+}$    &$\textbf{--}$          &  $\textbf{--}$ &   $\textbf{--}$& $\textbf{--}$\\
\SRUM & $\textbf{+}^*$              &$\textbf{--}$    &$\textbf{--}$  & $+$ only in \SRUM2 & $+$ only in \SRUM2&  $+$ only in \SRUM2\\
\OSPUM & $\textbf{--}$              &$\textbf{+}$    &$\textbf{+}$          & $\textbf{+}$ & close to peak&    $ \textbf{--}$\\
\DFUM & $\textbf{--}$              &$\textbf{+}$    &$\textbf{--}$          &$\textbf{--}$  &$\textbf{--}$  &  $\textbf{--}$\\
\bottomrule
\end{tabular}
\begin{tablenotes}
\item \emph{Notes:} * valid for all common prior models with $n=2$ and for generic priors for $n>2$ \citep[see][]{Schummer-Velez-2019}.
\end{tablenotes}
\end{threeparttable}
}
\caption{Predictions. $(+)$ The mechanism obtains the Uniform outcome with certainty for each possible common prior/agents have available the corresponding type of strategy; ($-$) there are ex-post equilibria that do not obtain the Uniform outcome (first column), or the mechanism does not possess the corresponding equilibrium/or type of strategy (second-sixth column). \SRUM2 is the second mover in \SRUM.\label{tab:EqPredictions}}
\end{table}

Let us refer to the first and second stages of \SRUM\ as \SRUM1 and \SRUM2, respectively. From an informational perspective, \SRUM1\ is a different problem than \DRUM. \SRUM1 requires the agent to understand the consequences of their choice given that the second agent will best respond under certainty. By contrast, in \DRUM, the agent is making a choice without information about the choice of the other agent. Reasoning in \SRUM1\ will require no guess about the behavior of the second mover, provided it maximizes the first mover's payoff. This first mover has an equilibrium set of actions that include---but do not single out---the agent's peak.\footnote{Note that reporting the peak at \SRUM1 may not be a best response for the first mover if \SRUM2 mover is not a payoff maximizer.} If an agent anticipates some noisy behavior by their opponent in \DRUM, introspection leads to the dominant strategy of peak reporting as the only payoff maximizing choice. 

The second benchmark is (weak) Dominant Strategy equilibrium. Except for the agent who moves at \SRUM1, all other players posses Dominant Strategies in our games. These strategies obtain the Uniform outcome in \DRUM, \OSPUM, and \DFUM.  

\begin{table}[t]
\centering
\begin{tabular}{cccccccccccc}
   $\theta_2$& $\leq10$ & 11 & 12 & 13 & 14 & 15 & 16 & 17 & 18 & 19 &20 \\\hline

 $u_1(4,\theta_2)$&  15 & 16 & 17 & 18 & 19 & 20 & 19 & 19 & 19 & 19&19 \\
   $u_1(5,\theta_2)$& 15 & 16 & 17 & 18 & 19 & 20 & 20 & 20 & 20 & 20&20 \\
  \hline
\end{tabular}
\caption{Contingent reasoning in Simultaneous Uniform Rationing game.}
\label{Tab:contingent-reasoning}
\end{table}

Together, Nash equilibrium and Dominant Strategy Equilibrium do not hint at a particular mechanism as being potentially superior in terms of its capacity to produce its intended outcomes. If dominant strategies, just for being dominant, are indeed focal, all the dominant strategy mechanisms should obtain the Uniform outcome. \SRUM\ lacks dominant strategies for the first mover. However, this player has a strong incentive to choose an action that induces the Uniform outcome when the second mover is payoff maximizer. Indeed, any perfect equilibrium strategy by \SRUM1 mover can be replaced by the truthful strategy and the resulting profile of strategies is again a perfect equilibrium \citep{Schummer-Velez-2019}. Thus, peak reporting for \SRUM1 is as simple as a dominant strategy from the point of view of strategic uncertainty: it is always a best response to any equilibrium action.\footnote{See \citet{Borgers-Li-2019-Eca} for a formalization of a related property.}

There is ample evidence of persistent weakly dominated behavior in some games that posses dominant strategies \citep[see][for a survey]{Velez-Brown-2019-SP}. This phenomenon has sparked great interest in identifying properties of dominant strategy mechanisms that promote the choice of dominant strategies. A key observation is that agents drop from a clock auction at their value at a higher rate than the frequency of truthful reports in second-price auctions \citep{Kegel-et-al-1987-Eca}.  \citet{Li-AER-17} proposed a theoretical explanation to this phenomenon based on a mechanism's requirements for contingent reasoning. \citet{Pycia-Troyan-2019} elaborate further, proposing the categorization of simplicity of mechanisms which we describe next.

\DRUM\ illustrates how identifying a dominant strategy may require contingent reasoning.  That is, discovering that reporting an agent's peak weakly dominates any other report requires that the agent conditions the comparison on a given report of the other agent.  Suppose that agent~$1$ has peak $5$. Table~\ref{Tab:contingent-reasoning} shows the different payoffs that are obtained by reports $4$ and $5$ as a function of the other agent's report. If the agent understands the structure of the game in detail, they know that (in a one-shot realization of the game) their actions do not influence the behavior of the other agent. Thus, reporting $5$ will always lead to a payoff that is no less than reporting $4$, and in some events the payoff of $5$ will be higher. If the agent has a coarse understanding of the game, i.e., can only grasp that reporting $4$ or $5$ both lead to payoffs $\{15,...,20\}$ in some events, but is not able to understand the logical relation between these events, reporting $5$ may not be identified as unambiguously better than $4$.

To formalize this idea and also articulate the notion of limited control of future actions, \citet{Pycia-Troyan-2019}  introduce the concept of \emph{$k$-step strategic collection of strategic plans} (for $k\geq0$): an action at each node in which an agent is called to move, and the $k$-consecutive-action paths (modulo termination) that follow the action. Differently from a standard strategy, these collections of strategic plans (indexed by the nodes where the agent is called to play) do no need to be dynamically consistent.

A $k$-step strategic plan is Simply Dominant at a given node in which the agent is called to play if the worst outcome reached by following the plan is no worse than the best outcome that can be obtained by choosing a different action from the one prescribed by the plan at the node.

Simple Dominance of a strategic plan captures the idea that identifying the plan as optimal requires no contingent reasoning. The agent only needs to know the range of payoffs they can obtain following some alternative action and not the exact relationship between them. For instance, in \DRUM , $k$-strategic plans (for any $k\geq 0$) coincide with standard actions at the only node in which agents are called to play. Even though $5$ is dominant for an agent with this peak, it is not $0$-step Simply Dominant: the worst payoff after reporting $5$, i.e., $15$, is worse than the best payoff after reporting $4$, i.e., $20$.

Since we can consider strategic plans of different length, we can also evaluate the level of confidence the agent has on their capacity to complete their action plans. 
A $k$-step Simply Dominant strategic plan, is intuitively simpler than a $(k+1)$-step Simply Dominant strategic plan. The shorter plan is better independently of the agent being able to complete the $(k+1)$-th action.

\begin{figure}[]
\centering
\includegraphics[width=0.7\textwidth]{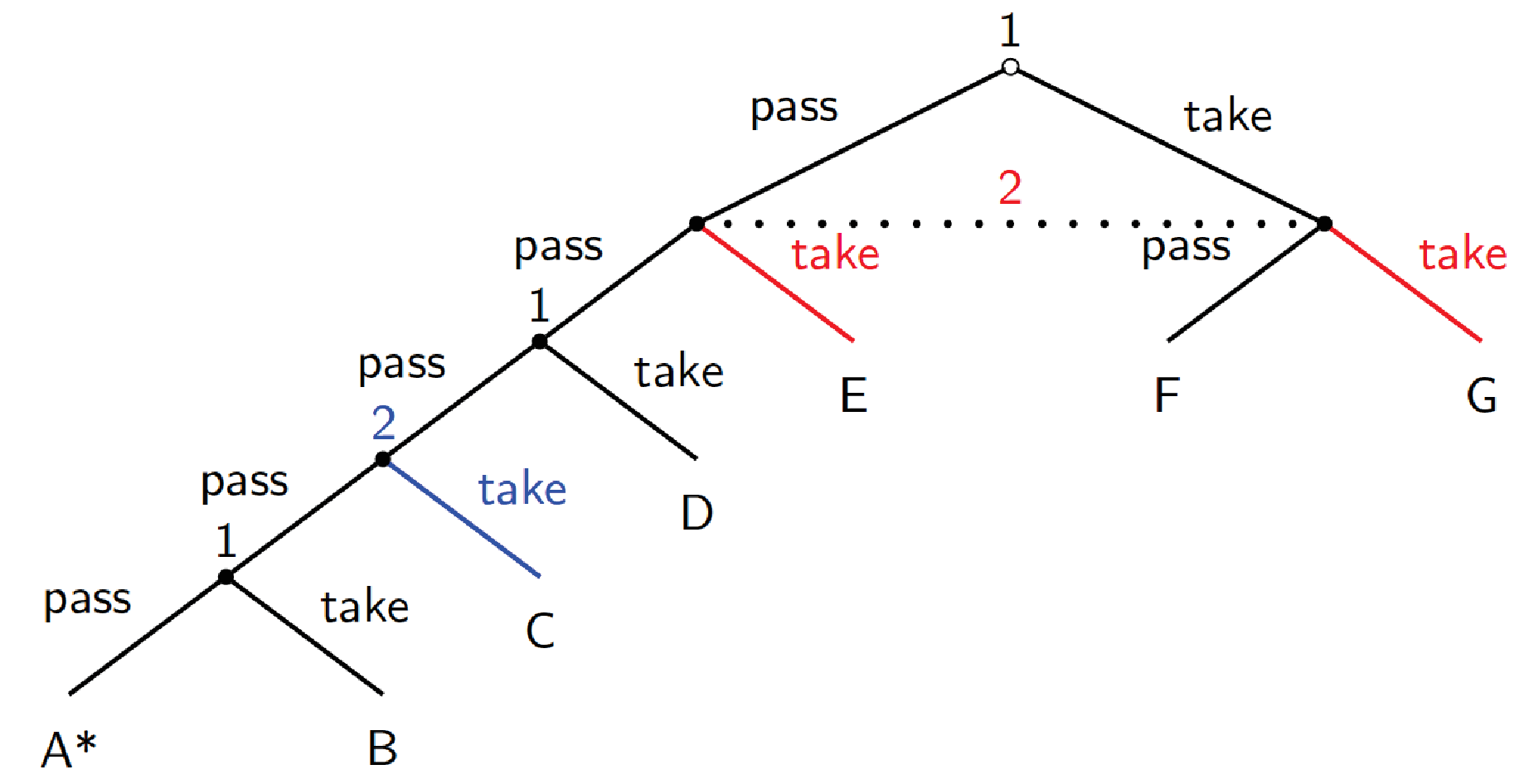}
\caption{An example game demonstrating the hierarchy of simplicity concepts involved with an agent identifying a dominant strategy. Suppose Player 1's most (least) preferred outcome is $A$ ($G$) and preferences are such that $C\succsim D, F$. A strategy of choosing ``pass'' at each node (PPP) is a dominant strategy. PPP is \emph{Obviously Dominant} if we add $E \succsim F$. Should either $D \succsim F$ or $B \succsim F$ also hold, pass at the initial node is part of a \emph{1-step Simply Dominant} path of play. It is part of a \emph{0-step Simply Dominant} path of play if both final conditions also hold. \label{fig:hierarchy}}
\end{figure}

Figure \ref{fig:hierarchy} provides an example in a different context to illustrate this hierarchy. 
Suppose Player 1's most (least) preferred outcome is $A$ ($G$) and preferences are such that $C\succsim D, F$. A strategy of choosing ``pass'' at each node (PPP) is a dominant strategy. Player 2 will either play ``take'' at the second node and 1 realizes $E$ rather than $G$, or 2 plays ``pass'' and the game continues with $A$ or $C$ eventually realized. %
Should we add $E \succsim F$ to the previous conditions, the three possible outcomes that can occur for 1 under PPP are at least as good as the two possible outcomes from playing ``take'' at the first node. The strategy PPP would be \emph{Obviously Dominant}.

If additionally either $D \succsim F$ or $B \succsim F$ hold, PPP is the equilibrium play of an agent whose intended plans of play are \emph{1-step Simply Dominant}. Under the former condition, at the initial node, 1 must only plan to ``pass'' and then to ``take'' at the next node to get $D$ and guarantee a better outcome than choosing ``take'' at  the initial node. For the latter case, 1 can plan to pass at the second node (and not plan further) knowing that outcomes $A$, $B$ or $C$ are all better than any outcome should ``take'' be played. If both conditions hold, any outcome after ``pass'' is played at the first node is at least as good as any outcome when ``take'' is played at that node, meaning no further planning is required to recognize an obviously dominant action, such case is \emph{0-step Simply Dominant}.

\cite{Pycia-Troyan-2019} characterize mechanisms using this heirarchy. Three categories are relevant:

\begin{itemize}
  \item Obviously Strategy-proof (\OSP): There is a $\infty$-step Simply Dominant strategic collection that obtains the desired outcomes. (Coincides with \citet{Li-AER-17}'s notion of simplicity.)
  \item One-Step Simple (\OSS): There is a $1$-step Simply Dominant strategic collection that obtains the desired outcomes.
  \item Strongly Obviously Strategy-proof (\SOSP): There is a $0$-step Simply Dominant strategic collection that obtains the desired outcomes.
\end{itemize}

\OSPUM\ is \OSP: Each agent moving towards the peak at each node where they are called to play is an $\infty$-step Simple Strategic Collection. Following the plan never gets the agent farther to the peak and deviating never gets the agent closer to the peak. \OSPUM\ is also \OSS: aach agent moving towards the peak at each node where they are called to play and leaving the game in the next node they are called to play never gets the agent farther to the peak and deviating never gets the agent closer to the peak. 

\OSPUM\ has no $0$-step Simply Dominant strategic collection, however. (Thus, it is not \SOSP.) For instance, when the agent is called to play close enough to their peak, moving towards the peak opens the possibility that the agent ``sleeps on the wheel'' and keeps moving towards the end of the interval and passes the peak. Suppose that the agent has peak $9$ and is called to play at the initial node of the game, i.e., to choose between $9$, $10$, and $11$. If the agent chooses $10$, their payoff is $19$. If they choose $9$ and do not opt out at the next nodes, if the other agent keeps playing, the agent may end up in $0$, which gives them a payoff of $10$. 

Some actions in \OSPUM\ do belong to a $0$-step Simply Dominant strategic plan. Indeed if the agent is called to play far enough from their peak, the worst that happens if they move towards their peak is never worse than the best that happens if they opt out. For instance, at the initial node an agent with peak $3$ who chooses $9$, will at worst stay at $10$ with a payoff of $13$, which is exactly the best that can happen if they opt out. (Note that if this agent moves to $9$ and sleeps on the wheel, ends up at $0$, which gives a payoff of $17$).

Finally, 
%
%
agents in \DFUM\ and \SRUM2 have informational advantages compared to the other mechanisms. Indeed, \SRUM2's choice is the simplest type of problem one can envision: the agent makes a determinate choice.\footnote{See \cite{martinez2019failures} for empirical evidence of benefits of this particular type of simplicity.} In \DRUM, \SRUM1, and \OSPUM, there is a non-deterministic component in the realization of outcomes. Ranking actions depends on the capacity of the agent to envision those outcomes following the different paths of actions. As discussed above, \OSPUM\ involves comparisons of payoffs that can be done with a coarser understanding of the game than \DRUM\ and \SRUM1. Behaving optimally in \SRUM2 is still simpler than following a $0$-step Simply Dominant strategic plan in \OSPUM. In the latter, payoffs depend on histories of play. In the former, payoffs are simply in front of the agent.

\DFUM\ approximates to some extent the simplicity of \SRUM2. It may provide information about the intended action of each player. Best responding to this intention is a determinate problem. If they have not chosen their peak, the player may want to revise their choices as the other agent revises theirs too. For this reason, \DFUM\ is not as simple as \SRUM2. The final performance of \DFUM\ depends on how this process evolves and whether this non-binding information is related to actual chosen strategies (i.e., not entirely cheap talk).

\subsection{Hypotheses}
Our theoretical benchmarks allow us to make the following hypotheses on Uniform outcomes.   Since peak-reporting strategies are still focal in \SRUM1, and \SRUM2 is an unambiguously simple problem, it is plausible that \SRUM\ outperforms \DRUM\ \citep{Schummer-Velez-2019}. Since  \OSPUM\ is \OSS\ and \DRUM\ is not \OSP , 
one can hypothesize that \OSPUM\ outperforms \DRUM\ \citep{Li-AER-17,Pycia-Troyan-2019}. Finally, note that the payoff relevant move in \DFUM\ is equivalent to \DRUM. However, \DFUM\ allows unsophisticated agents to converge on Uniform outcomes (with significant probability) without the need of any introspection about the game.\footnote{As a benchmark, we can calculate the limiting distributions of a myopic best response dynamics (revised at times determined by a Poisson process with positive arrival time). Even though outcomes different from the Uniform are persistent, the percentage of Uniform outcomes that obtain is uniformly high (96.8\%) across valuation structures we test. It is worth noting that these dynamics do not predict a uniformly high (across valuations) rate of dominant strategy play (two valuations obtain around 10\%, the other four around 54\%). See our working paper for details.} Thus, it is plausible that \DFUM\ also outperforms \DRUM .

\begin{innerhypothesis}[\textbf{M}]\rm 
For M in $\{\rm \SRUM,\OSPUM,\DFUM\}$, mechanism M produces a higher frequency of Uniform outcomes than \DRUM.\label{Hyp:(M)}
\end{innerhypothesis}


Under dominant strategy mechanisms \DRUM , peak reporting is a dominant strategy. Backwards induction (perfection) does not single out an agent's peak as the only rational choice at both stages of \SRUM. We expect peak reports are more frequent in \DRUM\ than in \SRUM.

\begin{innerhypothesis}\rm
Subjects report their peak with a higher frequency in \DRUM\ than at each stage of \SRUM.\label{Hyp:DS_DRU>DS_SRU}
\end{innerhypothesis}

\OSPUM\ achieves the Uniform outcome in Obviously Dominant Strategies in its subgame perfect equilibrium. \DRUM\ attains the Uniform outcome in Nash equilibrium under dominance. Due to the added simplicity, we expect the rate of Obvious Strategy paths, i.e., agents moving towards their peak whenever possible would be higher than the rate of Uniform outcomes in \DRUM .
\footnote{Note that observed frequencies of on-equilibrium-path play in \OSPUM\ are an upper bound of frequencies of on-equilibrium-path play.}

\begin{innerhypothesis}\rm
Whenever feasible, subjects move towards their peak in \OSPUM\ with a higher frequency than they truthfully report in \DRUM.\label{Hyp:DS_OSP>DS_DRU}
\end{innerhypothesis}

Finally, \citet{Pycia-Troyan-2019}'s characterization of simplicity allows us to differentiate simplicity of play at the node level in \OSPUM. It is important to note that should the mechanism move in a direction away from a player's peak---either going past the peak or the wrong direction from 10---to stop the mechanism at the current node is called for by a $0$-step Simply Dominant path. Taken together, these observations confirm that subjects who exhibit different paths of play with the  \OSPUM\ will have a different number of encounters with nodes where the dominant strategy path satisfies 0- or 1-step Simple Dominance. Thus, there are selection issues behind any direct apples-to-apples comparison of rates of dominant-strategy actions at these two types of nodes. Nonetheless, if such issues could be addressed, we would expect higher rates of actions consistent with the dominant strategy at simpler nodes under \citet{Pycia-Troyan-2019}'s characterization.

\begin{innerhypothesis}
\rm Actions that are called for by a $0$-step Simply Dominant strategic plan are played more often than actions that are only called for by a $1$-step Simply Dominant strategic plan.\label{Hyp:Py-T}
\end{innerhypothesis}

\section{Results}\label{Sec.Results}

\subsection{Outcome measures}\label{SEc:Outcome_measures}
We begin our analysis by ranking mechanisms based on their capacity to produce desirable outcomes. Out of all Hypotheses \ref{Hyp:(M)}-M, we find solid empirical support for \DFUM\ producing a significantly higher frequency of Uniform outcomes than \DRUM\ (i.e., Hypothesis \ref{Hyp:(M)}-\DFUM ), and modest evidence for \SRUM\ (i.e., Hypothesis \ref{Hyp:(M)}-\SRUM ).  

\begin{result}\label{Res:outcomes}\, In terms of outcomes,
\begin{enumerate}[label=\alph*.]
\item \DFUM\ performs unambiguously better than the other mechanisms. Independent of valuation structure, \DFUM\ achieves a higher rate of Uniform outcomes compared to any other mechanism tested. It also achieves the highest rate of efficient allocations, almost efficient allocations, and shares of earnings relative to efficient allocations. 
\item \SRUM\ generally achieves more Uniform outcomes and higher efficiency measures than \DRUM\  and \OSPUM, especially in cases where the Uniform allocation differs from equal division.
\end{enumerate}
\end{result}

\begin{table}[]
\centering
\begin{tabular}{ccccccc}
\toprule
\multicolumn{7}{c}{Panel A: rate of Uniform outcomes}\\
\midrule
&valuation & \DRUM\  & \SRUM\   & \OSPUM\           & \DFUM\   \\
  & 1             &0.788               &0.684               &0.696               &0.875\\
    &    2              &0.865               &0.632               &0.739               &0.958\\
      &  3              &0.269               &0.421               &0.196               &0.583\\
       & 4              &0.423               &0.395               &0.239               &0.625\\
       & 5              &0.212               &0.474               &0.261               &0.583\\
       & 6              &0.365               &0.632               &0.500               &0.792\\
&overall             &0.487               &0.539               &0.438               &0.736\\
\midrule
\multicolumn{7}{c}{Panel B: Rate of efficient outcomes}\\
\midrule
&valuation & \DRUM\  & \SRUM\   & \OSPUM\           & \DFUM\   \\
&1 &0.981                &0.974                &1.000                &0.958 \\
&2 &1.000              & 0.974               & 1.000               & 1.000  \\
&3 &0.269              & 0.421              & 0.196                &0.583   \\
&4 &0.538               & 0.447               & 0.283              & 0.708  \\
&5 &0.231                &0.579               &0.304  &              0.750 \\
&6 &0.365               & 0.632               &0.500  &              0.792 \\
&overall &0.564                &0.671              & 0.547               &0.799   \\
\midrule
\multicolumn{7}{c}{Panel C: rate of within 1 of efficient outcome}\\
\midrule
&valuation & \DRUM\  & \SRUM\   & \OSPUM\           & \DFUM\   \\
&1 &0.981              & 0.974               &1.000               &1.000    \\
&2 &1.000               &0.974               &1.000                &1.000   \\
&3 &0.308               &0.526               &0.283               &0.667    \\
&4 &0.596               &0.500                &0.304               & 0.750  \\
&5 &0.308               &0.632               &0.326              & 0.875    \\
&6 &0.962               &1.000              & 0.935                & 1.000  \\
&overall &0.692               &0.768               &0.641               &0.882    \\
\midrule
\multicolumn{7}{c}{Panel D: share of efficient outcome earnings}\\
\midrule
&valuation & \DRUM\  & \SRUM\   & \OSPUM\           & \DFUM\   \\
&1 &0.996                &0.994                &1.000                &0.997 \\
&2 &1.000                &0.993               &1.000               &1.000   \\
&3 &0.838               &0.855              & 0.784               &0.902    \\
&4 &0.929              & 0.905              & 0.880               &0.956    \\
&5 &0.849               &0.903               &  0.818              & 0.961  \\
&6 &0.964               &0.982               &0.972               &0.990    \\
&overall &0.930               &0.939               & 0.909               & 0.968  \\
\bottomrule
\end{tabular}
\caption{\label{overall}Overall performance measures for each of the four mechanisms.}
\end{table}

\begin{table}[]
\centering
\begin{threeparttable}\footnotesize
\begin{tabular}{lcccc}
&(1)&(2)&(3)&(4)\\
&\begin{tabular}{c}Uniform\\ rule \\outcome\end{tabular}&\begin{tabular}{c}efficient\\ outcome\end{tabular}&\begin{tabular}{c}within-1\\ of efficient\\ outcome\end{tabular}&\begin{tabular}{c}share of\\ efficiency\end{tabular}\\\hline
second half& 0.050 & 0.023 & 0.040* & 0.007 \\
 & (0.028) & (0.016) & (0.022) & (0.005) \\
CFU& 0.249*** & 0.235*** & 0.190*** & 0.038*** \\
 & (0.050) & (0.057) & (0.034) & (0.010) \\
OSPU& -0.049 & -0.017 & -0.051 & -0.021 \\
 & (0.048) & (0.053) & (0.051) & (0.013) \\
SRU& 0.052* & 0.107*** & 0.075** & 0.009 \\
 & (0.028) & (0.028) & (0.033) & (0.006) \\
\hline
observation level&decision-pair&decision-pair&decision-pair&decision-pair\\
observations&960&960&960&960\\
 r-squared & 0.178 & 0.337 & 0.376 & 0.365 \\\hline
\end{tabular}
\begin{tablenotes}
\item \emph{Notes:} \DRUM\ is baseline (omitted) treatment. Valuation dummy variables are also included in all regressions. All regression models use cluster-robust standard errors at the session level.
\end{tablenotes}
\end{threeparttable}
\caption{Regression analysis of pair-level outcomes by mechanism\label{reg:pair}}
\end{table}

Table \ref{overall} provides four different performance measures for each of the four mechanisms separated by valuation and overall: Frequencies of Uniform outcome, efficient, and near efficient outcomes, and share of efficient outcome earnings realized. In our environment an allocation is efficient if both agents receive an allotment on the same side of their peak \citep{Sprumont-1991}. Since agents' utility has the same slope on both sides of their peak, the summation of their utilities is constant on all efficient allocations. Thus, this last statistic is a meaningful measure of performance, as well as the distribution of efficiency losses (see below).

The most striking result is that \DFUM\ achieves the highest performance over each measure across every valuation. Table~\ref{reg:pair} provides corresponding regression results. The \DFUM\ achieves roughly 25 percentage points higher rates of Uniform and efficient outcomes, a nearly 20 percentage point higher rate of a near efficient outcomes,\footnote{``Near efficiency'' refers to outcomes within 1 point of the fully efficient outcome. For example, an allocation of (5, 15) would be near efficient if the efficient outcome were (4, 16).} and a 4 percentage point higher share of maximum possible earnings than  the baseline \DRUM\ mechanism ($p<0.001$, all four comparisons). These results are not consistent with dominant strategy equilibrium and perfect equilibrium predictions, which suggest similar performance should be observed across all mechanisms (see Sec~\ref{Sec:Eq-predictions} and Table~\ref{tab:EqPredictions}).

\begin{figure}[t]
\centering
\begin{pspicture}(-.5,-2.5)(12.5,9)
\psframe(1,-1.5)(12.5,-1)
\rput[l](1.5,-1.25){$\mbox{\footnotesize\DRUM\ }$}
\rput[l](4.125,-1.25){$\mbox{\footnotesize\SRUM\ }$}
\rput[l](6.75,-1.25){$\mbox{\footnotesize\OSPUM\ }$}
\rput[l](9.375,-1.25){$\mbox{\footnotesize\DFUM\ }$}
\psline[linecolor=gray,linestyle=solid](2.625,-1.25)(3.625,-1.25)	\psline[linecolor=blue,linestyle=dashed](5.25,-1.25)(6.25,-1.25)	\psline[linewidth=2pt,dotsep=.5pt,linecolor=red,linestyle=dotted](7.875,-1.25)(8.875,-1.25)	\psline[linecolor=black,linestyle=solid](10.5,-1.25)(11.5,-1.25)
\psaxes[ticks=y,yticksize=0 5,ysubticks=5,
  subticksize=1,tickcolor=black!20,subtickcolor=black!30,
  subticklinestyle=dashed,Dy=0.25,dy=.9cm,Ox=-24,dx=1.249cm,Dx=6](1,5)(6,8.6)
\psaxes[labels=none,ticks=x,Ox=-24,dx=1.249cm,Dx=6](1,5)(6,8.6)
\psframe*[linecolor=gray](1,8.6)(6,9)
\psline(6,5)(6,9)(1,9)(1,8.6)(6,8.6)
\rput[c](3.5,8.8){\footnotesize Val 3 $\mbox{\footnotesize $(4,16)$}$}
\psline[linecolor=gray,linestyle=solid](1,5)(1.41666666666667,5)	\psline[linecolor=blue,linestyle=dashed](1,5)(1.41666666666667,5)	\psline[linewidth=2pt,dotsep=.5pt,linecolor=red,linestyle=dotted](1,5)(1.41666666666667,5)	\psline[linecolor=black,linestyle=solid](1,5)(1.41666666666667,5.15)
\psline[linecolor=gray,linestyle=solid](1.41666666666667,5)(1.83333333333333,5)	\psline[linecolor=blue,linestyle=dashed](1.41666666666667,5)(1.83333333333333,5.09473684210526)	\psline[linewidth=2pt,dotsep=.5pt,linecolor=red,linestyle=dotted](1.41666666666667,5)(1.83333333333333,5.07826086956522)	\psline[linecolor=black,linestyle=solid](1.41666666666667,5.15)(1.83333333333333,5.15)
\psline[linecolor=gray,linestyle=solid](1.83333333333333,5)(2.25,5)	\psline[linecolor=blue,linestyle=dashed](1.83333333333333,5.09473684210526)(2.25,5.09473684210526)	\psline[linewidth=2pt,dotsep=.5pt,linecolor=red,linestyle=dotted](1.83333333333333,5.07826086956522)(2.25,5.15652173913043)	\psline[linecolor=black,linestyle=solid](1.83333333333333,5.15)(2.25,5.15)
\psline[linecolor=gray,linestyle=solid](2.25,5)(2.66666666666667,5)	\psline[linecolor=blue,linestyle=dashed](2.25,5.09473684210526)(2.66666666666667,5.18947368421053)	\psline[linewidth=2pt,dotsep=.5pt,linecolor=red,linestyle=dotted](2.25,5.15652173913043)(2.66666666666667,5.23478260869565)	\psline[linecolor=black,linestyle=solid](2.25,5.15)(2.66666666666667,5.15)
\psline[linecolor=gray,linestyle=solid](2.66666666666667,5)(3.08333333333333,5)	\psline[linecolor=blue,linestyle=dashed](2.66666666666667,5.18947368421053)(3.08333333333333,5.18947368421053)	\psline[linewidth=2pt,dotsep=.5pt,linecolor=red,linestyle=dotted](2.66666666666667,5.23478260869565)(3.08333333333333,5.31304347826087)	\psline[linecolor=black,linestyle=solid](2.66666666666667,5.15)(3.08333333333333,5.15)
\psline[linecolor=gray,linestyle=solid](3.08333333333333,5)(3.5,6.10769230769231)	\psline[linecolor=blue,linestyle=dashed](3.08333333333333,5.18947368421053)(3.5,6.51578947368421)	\psline[linewidth=2pt,dotsep=.5pt,linecolor=red,linestyle=dotted](3.08333333333333,5.31304347826087)(3.5,7.19130434782609)	\psline[linecolor=black,linestyle=solid](3.08333333333333,5.15)(3.5,5.75)
\psline[linecolor=gray,linestyle=solid](3.5,6.10769230769231)(3.91666666666667,6.38461538461539)	\psline[linecolor=blue,linestyle=dashed](3.5,6.51578947368421)(3.91666666666667,6.51578947368421)	\psline[linewidth=2pt,dotsep=.5pt,linecolor=red,linestyle=dotted](3.5,7.19130434782609)(3.91666666666667,7.34782608695652)	\psline[linecolor=black,linestyle=solid](3.5,5.75)(3.91666666666667,5.75)
\psline[linecolor=gray,linestyle=solid](3.91666666666667,6.38461538461539)(4.33333333333333,6.86923076923077)	\psline[linecolor=blue,linestyle=dashed](3.91666666666667,6.51578947368421)(4.33333333333333,6.51578947368421)	\psline[linewidth=2pt,dotsep=.5pt,linecolor=red,linestyle=dotted](3.91666666666667,7.34782608695652)(4.33333333333333,7.34782608695652)	\psline[linecolor=black,linestyle=solid](3.91666666666667,5.75)(4.33333333333333,5.9)
\psline[linecolor=gray,linestyle=solid](4.33333333333333,6.86923076923077)(4.75,7.14615384615385)	\psline[linecolor=blue,linestyle=dashed](4.33333333333333,6.51578947368421)(4.75,6.51578947368421)	\psline[linewidth=2pt,dotsep=.5pt,linecolor=red,linestyle=dotted](4.33333333333333,7.34782608695652)(4.75,7.42608695652174)	\psline[linecolor=black,linestyle=solid](4.33333333333333,5.9)(4.75,6.2)
\psline[linecolor=gray,linestyle=solid](4.75,7.14615384615385)(5.16666666666667,7.49230769230769)	\psline[linecolor=blue,linestyle=dashed](4.75,6.51578947368421)(5.16666666666667,6.70526315789474)	\psline[linewidth=2pt,dotsep=.5pt,linecolor=red,linestyle=dotted](4.75,7.42608695652174)(5.16666666666667,7.58260869565217)	\psline[linecolor=black,linestyle=solid](4.75,6.2)(5.16666666666667,6.2)
\psline[linecolor=gray,linestyle=solid](5.16666666666667,7.49230769230769)(5.58333333333333,7.63076923076923)	\psline[linecolor=blue,linestyle=dashed](5.16666666666667,6.70526315789474)(5.58333333333333,7.08421052631579)	\psline[linewidth=2pt,dotsep=.5pt,linecolor=red,linestyle=dotted](5.16666666666667,7.58260869565217)(5.58333333333333,7.89565217391304)	\psline[linecolor=black,linestyle=solid](5.16666666666667,6.2)(5.58333333333333,6.5)
\psline[linecolor=gray,linestyle=solid](5.58333333333333,7.63076923076923)(6,8.6)	\psline[linecolor=blue,linestyle=dashed](5.58333333333333,7.08421052631579)(6,8.6)	\psline[linewidth=2pt,dotsep=.5pt,linecolor=red,linestyle=dotted](5.58333333333333,7.89565217391304)(6,8.6)	\psline[linecolor=black,linestyle=solid](5.58333333333333,6.5)(6,8.6)
%
%
%
\psaxes[ticks=y,yticksize=0 5,ysubticks=5,
  subticksize=1,tickcolor=black!20,subtickcolor=black!30,
  subticklinestyle=dashed,Dy=0.25,dy=.9cm,Ox=-20,dx=1cm,Dx=4](7.5,5)(12.5,8.6)
\psaxes[labels=none,ticks=x,Ox=-20,dx=1cm,Dx=4](7.5,5)(12.5,8.6)
\psframe*[linecolor=gray](7.5,8.6)(12.5,9)
\psline(12.5,5)(12.5,9)(7.5,9)(7.5,8.6)(12.5,8.6)
\rput[c](10,8.8){\footnotesize Val 4 $\mbox{\footnotesize $(3,13)$}$}
\psline[linecolor=gray,linestyle=solid](7.5,5)(8,5)	\psline[linecolor=blue,linestyle=dashed](7.5,5)(8,5.09473684210526)	\psline[linewidth=2pt,dotsep=.5pt,linecolor=red,linestyle=dotted](7.5,5)(8,5)	\psline[linecolor=black,linestyle=solid](7.5,5)(8,5)
\psline[linecolor=gray,linestyle=solid](8,5)(8.5,5)	\psline[linecolor=blue,linestyle=dashed](8,5.09473684210526)(8.5,5.09473684210526)	\psline[linewidth=2pt,dotsep=.5pt,linecolor=red,linestyle=dotted](8,5)(8.5,5)	\psline[linecolor=black,linestyle=solid](8,5)(8.5,5)
\psline[linecolor=gray,linestyle=solid](8.5,5)(9,5)	\psline[linecolor=blue,linestyle=dashed](8.5,5.09473684210526)(9,5.09473684210526)	\psline[linewidth=2pt,dotsep=.5pt,linecolor=red,linestyle=dotted](8.5,5)(9,5)	\psline[linecolor=black,linestyle=solid](8.5,5)(9,5)
\psline[linecolor=gray,linestyle=solid](9,5)(9.5,5)	\psline[linecolor=blue,linestyle=dashed](9,5.09473684210526)(9.5,5.09473684210526)	\psline[linewidth=2pt,dotsep=.5pt,linecolor=red,linestyle=dotted](9,5)(9.5,5)	\psline[linecolor=black,linestyle=solid](9,5)(9.5,5)
\psline[linecolor=gray,linestyle=solid](9.5,5)(10,5)	\psline[linecolor=blue,linestyle=dashed](9.5,5.09473684210526)(10,5.09473684210526)	\psline[linewidth=2pt,dotsep=.5pt,linecolor=red,linestyle=dotted](9.5,5)(10,5)	\psline[linecolor=black,linestyle=solid](9.5,5)(10,5)
\psline[linecolor=gray,linestyle=solid](10,5)(10.5,5.06923076923077)	\psline[linecolor=blue,linestyle=dashed](10,5.09473684210526)(10.5,5.09473684210526)	\psline[linewidth=2pt,dotsep=.5pt,linecolor=red,linestyle=dotted](10,5)(10.5,5.23478260869565)	\psline[linecolor=black,linestyle=solid](10,5)(10.5,5)
\psline[linecolor=gray,linestyle=solid](10.5,5.06923076923077)(11,6.38461538461539)	\psline[linecolor=blue,linestyle=dashed](10.5,5.09473684210526)(11,6.61052631578947)	\psline[linewidth=2pt,dotsep=.5pt,linecolor=red,linestyle=dotted](10.5,5.23478260869565)(11,7.42608695652174)	\psline[linecolor=black,linestyle=solid](10.5,5)(11,5.9)
\psline[linecolor=gray,linestyle=solid](11,6.38461538461539)(11.5,6.45384615384615)	\psline[linecolor=blue,linestyle=dashed](11,6.61052631578947)(11.5,6.8)	\psline[linewidth=2pt,dotsep=.5pt,linecolor=red,linestyle=dotted](11,7.42608695652174)(11.5,7.50434782608696)	\psline[linecolor=black,linestyle=solid](11,5.9)(11.5,5.9)
\psline[linecolor=gray,linestyle=solid](11.5,6.45384615384615)(12,6.66153846153846)	\psline[linecolor=blue,linestyle=dashed](11.5,6.8)(12,6.98947368421053)	\psline[linewidth=2pt,dotsep=.5pt,linecolor=red,linestyle=dotted](11.5,7.50434782608696)(12,7.58260869565217)	\psline[linecolor=black,linestyle=solid](11.5,5.9)(12,6.05)
\psline[linecolor=gray,linestyle=solid](12,6.66153846153846)(12.5,8.6)	\psline[linecolor=blue,linestyle=dashed](12,6.98947368421053)(12.5,8.6)	\psline[linewidth=2pt,dotsep=.5pt,linecolor=red,linestyle=dotted](12,7.58260869565217)(12.5,8.6)	\psline[linecolor=black,linestyle=solid](12,6.05)(12.5,8.6)
%
%
\psaxes[ticks=y,yticksize=0 5,ysubticks=5,
  subticksize=1,tickcolor=black!20,subtickcolor=black!30,
  subticklinestyle=dashed,Dy=0.25,dy=.9cm,Ox=-16,dx=1.249cm,Dx=4](1,0)(6,3.6)
\psaxes[labels=none,ticks=x,Ox=-16,dx=1.249cm,Dx=4](1,0)(6,3.6)
\psframe*[linecolor=gray](1,3.6)(6,4)
\psline(6,0)(6,4)(1,4)(1,3.6)(6,3.6)
\rput[c](3.5,3.8){\footnotesize Val 5 $\mbox{\footnotesize $(5,17)$}$}
\psline[linecolor=gray,linestyle=solid](1,0)(1.625,0)	\psline[linecolor=blue,linestyle=dashed](1,0)(1.625,0)	\psline[linewidth=2pt,dotsep=.5pt,linecolor=red,linestyle=dotted](1,0)(1.625,0.156521739130435)	\psline[linecolor=black,linestyle=solid](1,0)(1.625,0)
\psline[linecolor=gray,linestyle=solid](1.625,0)(2.25,0)	\psline[linecolor=blue,linestyle=dashed](1.625,0)(2.25,0.0947368421052632)	\psline[linewidth=2pt,dotsep=.5pt,linecolor=red,linestyle=dotted](1.625,0.156521739130435)(2.25,0.547826086956522)	\psline[linecolor=black,linestyle=solid](1.625,0)(2.25,0)
\psline[linecolor=gray,linestyle=solid](2.25,0)(2.875,1.45384615384615)	\psline[linecolor=blue,linestyle=dashed](2.25,0.0947368421052632)(2.875,1.04210526315789)	\psline[linewidth=2pt,dotsep=.5pt,linecolor=red,linestyle=dotted](2.25,0.547826086956522)(2.875,2.2695652173913)	\psline[linecolor=black,linestyle=solid](2.25,0)(2.875,0.45)
\psline[linecolor=gray,linestyle=solid](2.875,1.45384615384615)(3.5,1.59230769230769)	\psline[linecolor=blue,linestyle=dashed](2.875,1.04210526315789)(3.5,1.32631578947368)	\psline[linewidth=2pt,dotsep=.5pt,linecolor=red,linestyle=dotted](2.875,2.2695652173913)(3.5,2.2695652173913)	\psline[linecolor=black,linestyle=solid](2.875,0.45)(3.5,0.45)
\psline[linecolor=gray,linestyle=solid](3.5,1.59230769230769)(4.125,2.00769230769231)	\psline[linecolor=blue,linestyle=dashed](3.5,1.32631578947368)(4.125,1.32631578947368)	\psline[linewidth=2pt,dotsep=.5pt,linecolor=red,linestyle=dotted](3.5,2.2695652173913)(4.125,2.2695652173913)	\psline[linecolor=black,linestyle=solid](3.5,0.45)(4.125,0.45)
\psline[linecolor=gray,linestyle=solid](4.125,2.00769230769231)(4.75,2.49230769230769)	\psline[linecolor=blue,linestyle=dashed](4.125,1.32631578947368)(4.75,1.32631578947368)	\psline[linewidth=2pt,dotsep=.5pt,linecolor=red,linestyle=dotted](4.125,2.2695652173913)(4.75,2.42608695652174)	\psline[linecolor=black,linestyle=solid](4.125,0.45)(4.75,0.45)
\psline[linecolor=gray,linestyle=solid](4.75,2.49230769230769)(5.375,2.76923076923077)	\psline[linecolor=blue,linestyle=dashed](4.75,1.32631578947368)(5.375,1.51578947368421)	\psline[linewidth=2pt,dotsep=.5pt,linecolor=red,linestyle=dotted](4.75,2.42608695652174)(5.375,2.50434782608696)	\psline[linecolor=black,linestyle=solid](4.75,0.45)(5.375,0.9)
\psline[linecolor=gray,linestyle=solid](5.375,2.76923076923077)(6,3.6)	\psline[linecolor=blue,linestyle=dashed](5.375,1.51578947368421)(6,3.6)	\psline[linewidth=2pt,dotsep=.5pt,linecolor=red,linestyle=dotted](5.375,2.50434782608696)(6,3.6)	\psline[linecolor=black,linestyle=solid](5.375,0.9)(6,3.6)
%
%
\psaxes[ticks=y,yticksize=0 5,ysubticks=5,
  subticksize=1,tickcolor=black!20,subtickcolor=black!30,
  subticklinestyle=dashed,Dy=0.25,dy=.9cm,Ox=-10,dx=1cm,Dx=2](7.5,0)(12.5,3.6)
\psaxes[labels=none,ticks=x,Ox=-10,dx=1cm,Dx=2](7.5,0)(12.5,3.6)
\psframe*[linecolor=gray](7.5,3.6)(12.5,4)
\psline(12.5,0)(12.5,4)(7.5,4)(7.5,3.6)(7.5,3.6)
\rput[c](10,3.8){\footnotesize Val 6 $\mbox{\footnotesize $(9,11)$}$}
\psline[linecolor=gray,linestyle=solid](7.5,0)(8.5,0.0692307692307692)	\psline[linecolor=blue,linestyle=dashed](7.5,0)(8.5,0)	\psline[linewidth=2pt,dotsep=.5pt,linecolor=red,linestyle=dotted](7.5,0)(8.5,0)	\psline[linecolor=black,linestyle=solid](7.5,0)(8.5,0)
\psline[linecolor=gray,linestyle=solid](8.5,0.0692307692307692)(9.5,0.0692307692307692)	\psline[linecolor=blue,linestyle=dashed](8.5,0)(9.5,0)	\psline[linewidth=2pt,dotsep=.5pt,linecolor=red,linestyle=dotted](8.5,0)(9.5,0)	\psline[linecolor=black,linestyle=solid](8.5,0)(9.5,0)
\psline[linecolor=gray,linestyle=solid](9.5,0.0692307692307692)(10.5,0.138461538461538)	\psline[linecolor=blue,linestyle=dashed](9.5,0)(10.5,0)	\psline[linewidth=2pt,dotsep=.5pt,linecolor=red,linestyle=dotted](9.5,0)(10.5,0.234782608695652)	\psline[linecolor=black,linestyle=solid](9.5,0)(10.5,0)
\psline[linecolor=gray,linestyle=solid](10.5,0.138461538461538)(11.5,2.28461538461538)	\psline[linecolor=blue,linestyle=dashed](10.5,0)(11.5,1.32631578947368)	\psline[linewidth=2pt,dotsep=.5pt,linecolor=red,linestyle=dotted](10.5,0.234782608695652)(11.5,1.8)	\psline[linecolor=black,linestyle=solid](10.5,0)(11.5,0.75)
\psline[linecolor=gray,linestyle=solid](11.5,2.28461538461538)(12.5,3.6)	\psline[linecolor=blue,linestyle=dashed](11.5,1.32631578947368)(12.5,3.6)	\psline[linewidth=2pt,dotsep=.5pt,linecolor=red,linestyle=dotted](11.5,1.8)(12.5,3.6)	\psline[linecolor=black,linestyle=solid](11.5,0.75)(12.5,3.6)
\rput[c]{90}(-.25,4.5){Empirical CDF}
\rput[c](6.75,-2){Utility loss, per group, compared to efficient allocation.}
\end{pspicture}
\caption{Stylized empirical CDF of utility loss, per group, compared to efficient allocation. Realized utility losses belong in $\{-24,-22,....,-2,0\}$. The lines in the graph join the values of the CDF in the domain of utility losses. Thus the values of the CDF can only be read from the graph for the realized utility losses. The lines allow an easy visual comparison of the distributions.}
\label{Fig:CDF-effloss}
\end{figure}
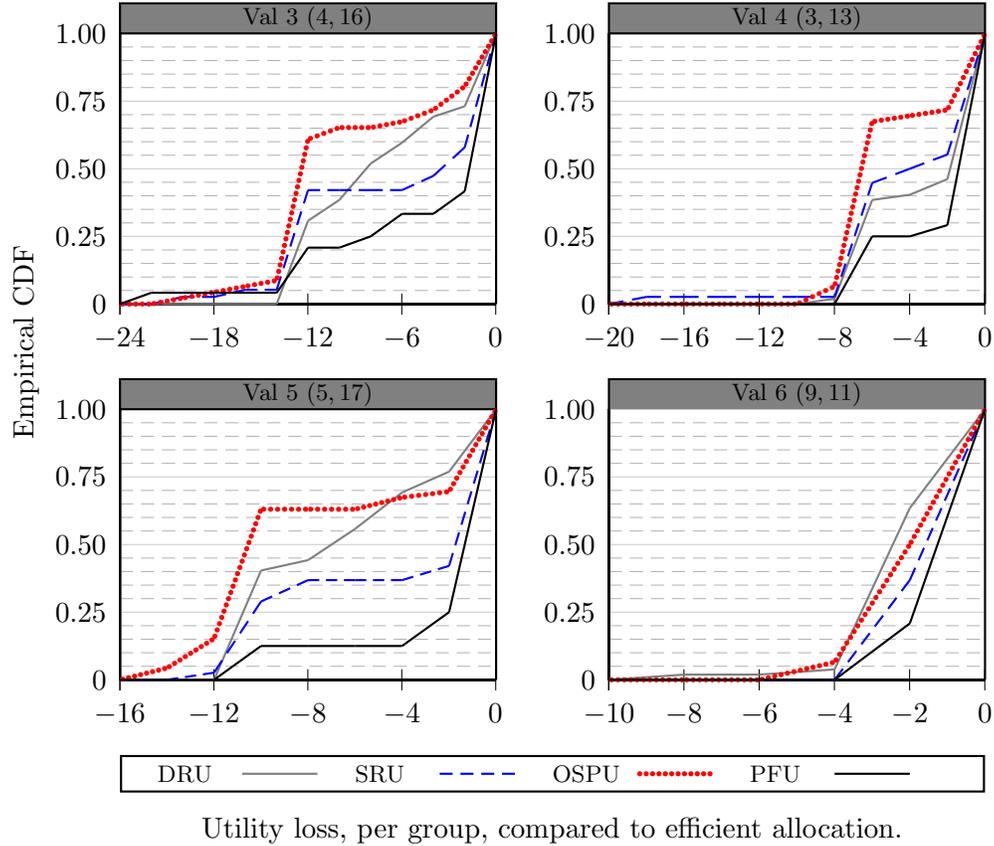

Figure~\ref{Fig:CDF-effloss} provides a granular robustness check on these result at the distribution level. It displays the cumulative distributions of utility loss, per group, compared to the efficient assignment for valuations where the Uniform allocation is not equal division (all mechanisms are essentially efficient in the other two valuation structures; see Table~\ref{overall}). With one exception for a low probability event in valuation $3$, $(4,16)$, for each level of efficiency loss $x$, the cumulative frequency of outcomes with a loss at least $x$ is lower for \DFUM\  than for each of the other mechanisms.

These results are robust to changes in valuation which clearly affect the realization of Uniform outcomes and other measures. The first two valuations have equal division, $(10,10)$, as Uniform outcome; the other valuations have a Uniform outcome which differs from equal division. Even though, on average, subjects encounter valuations 1 and 2 with less experience than the other valuations, these valuations achieve substantially higher measures of efficiency  (Table~\ref{overall}). The valuation dummy variables are jointly significantly different in Table~\ref{reg:pair} ($p<0.01$ all four specifications, not shown), even if we exclude all \OSPUM\ observations ($p<0.01$, all four specifications, regression and tests not shown). We further examine valuation differences at the end of Result \ref{Res:individual}.

While \SRUM\ also falls short of \DFUM\ in performance ($p<0.001$ all four measures, not shown), it is a clear second among the four mechanisms. For the first three measures, it outperforms \DRUM\ by 5--10 percentage points overall.  
The preceding results confirm Hypothesis~\ref{Hyp:(M)}-\SRUM\ which stated \SRUM\ would outperform \DRUM\ in terms of Uniform outcomes achieved. 

While \OSPUM\ does not differ significantly in performance from \DRUM\ in any of the four regressions in Table~\ref{reg:pair}, the direction of the performance is always negative. This difference is alarmingly inconsistent with Hypothesis~\ref{Hyp:(M)}-\OSPUM\ which would predict significantly positive relative performance. The performance differences between these two mechanisms further depends on the valuation pair. 
Table \ref{reg:pair-noeq} provides an additional interaction term for \OSPUM\ and valuations 3--6, the valuations that require the game to move beyond the initial stage to achieve the desired outcome. 
Relative to \DRUM , in valuations 1--2 \OSPUM\ significantly outperforms \DRUM\ by 3 of the 4 measures. In the later valuations it underperforms, though the difference is only marginally significant in one of the four measures: percentage of optimal earnings realized ($0.015+(-0.053)=-0.038$, $p<0.10$).

To summarize, outcomes measures single out \DFUM\ as the best performing mechanism among those tested. An important consequence is that there is little evidence to support hypotheses based on equilibrium, dominant strategies, or \OSP-related concepts of simplicity for predicting outcomes in rationing games. In the next results, we will see that there is some evidence that these concepts do predict individual behavior, but crucial differences across mechanisms prevent these individual differences from affecting outcomes. 

\subsection{Individual measures: peak reporting and best response}

\begin{result}\label{Res:individual} For the specific strategy profile associated with each agent reporting their true peak: 
\begin{enumerate}[label=\alph*.]
\item The rates of playing actions at nodes consistent with peak reporting is highest for \OSPUM . This value is somewhat reduced when we focus on the entirety of path of play.
\item Rates of best-response do not match rates of reporting true peaks. Subjects in \DFUM\ and \SRUM2\ have the highest rates of best response, nearly 90\%.
\item Uniform random play can predict many of the comparative statics across valuations, especially for valuations 1 and 2. Nonetheless, across all mechanisms pairs of subjects disproportionately play equilibrium profiles.
\end{enumerate}
\end{result}

The mechanisms we examine are designed to achieve the Uniform outcome based on the assumption of subjects truthfully reporting their peaks.
Our experiment reveals that the exact way in which the Uniform outcome is achieved is not necessarily reliant on truthful play and involves a very nuanced relationship involving many factors. Table \ref{tab:indy1}, Panel A provides summary rates of truthfully reporting one's peak under \DRUM, \SRUM, and \DFUM\ and of moving towards one's peak under \OSPUM\ across valuations. The asymmetric \SRUM\ is separated by first and second movers. Table \ref{reg:indy}, specifications (1) and (2) provide the corresponding results in regression form. 

A subject under \OSPUM\ is twenty-seven percentage points more likely to play an action consistent with a truthful report \emph{at a single node} than in the \DRUM\ mechanism---consistent with Hypothesis \ref{Hyp:DS_OSP>DS_DRU}---and 11.5 points more likely than \DFUM\ ($p<0.01$, both measures). If we consider the entirety of path of play, \OSPUM\ still has a higher rate than \DRUM\ and \SRUM2, but is no longer higher than \DFUM . Crucially, the performance is lowest over the last four valuations where moving towards one's peak requires actions at more than one node. 

Inconsistent with Hypothesis \ref{Hyp:DS_DRU>DS_SRU}, subjects in the \SRUM1, \SRUM2, and \DRUM  all report their peak at statistically identical rates. Result \ref{res:SRUM} examines this relationship further. Though not tied to a specific hypothesis, peak reporting rates are 15.1 points higher under \DFUM\ than \DRUM\ ($p<0.01$).

\begin{table}[]
\centering
\begin{tabular}{ccccccc}
\toprule
\multicolumn{7}{c}{Panel A: rate of actions/strategies consistent with}\\
\multicolumn{7}{c}{reporting/moving towards true peak}\\
\midrule
valuation & $\DRUM^a$ & \SRUM1$^b$ & \SRUM$^b$ & \multicolumn{2}{c}{\begin{tabular}{ll}\multicolumn{2}{c}{\OSPUM\ }\end{tabular}}          & $\DFUM^a$  \\
 & & & & Path-wise$^c$&Node-wise$^d$          &
\\
1             &0.240               &0.368                &0.158               &0.413&0.511	& 0.313\\
        2              &0.269               &0.158               &0.237                &0.533&0.571& 0.333\\
        3              &0.452               &0.421               &0.368                &0.391&0.745	& 0.625\\
        4              &0.385               &0.342               &0.368                &0.511&0.732	& 0.479\\
        5              &0.221                &0.263               &0.447               &0.478&0.777	& 0.479\\
        6              &0.548               &0.526               &0.632              &0.413&0.604 	& 0.792\\
overall             &0.353               &  0.346                &0.368               &0.457&0.678	& 0.503\\
\midrule
\multicolumn{7}{c}{Panel B: rate of actions/strategies consistent with playing best response}\\
\midrule
valuation & \DRUM\  & \SRUM1$^f$ & \SRUM2$^g$ & \multicolumn{2}{c}{\begin{tabular}{ll}\multicolumn{2}{c}{\OSPUM\ }\end{tabular}}          & \DFUM\   \\
 & & & & Path-wise&Node-wise$^g$          &
\\
1             &0.769               &0.789                &0.763               &0.663&0.640	& 0.917\\
        2              &0.875               &0.684               &0.789                &0.750&0.699& 0.958\\
        3              &0.635               &0.421               &0.868                &0.522&0.786	& 0.792\\
        4              &0.702               &0.553               &0.816                &0.652&0.799	& 0.813\\
        5              &0.615                &0.500               &0.947               &0.565&0.807	& 0.813\\
        6              &0.683               &0.684               &0.947               &0.565&0.701 	& 0.958\\
overall             &0.713               &  0.605                &0.855               &0.620&0.751	& 0.875\\
\bottomrule
\end{tabular}
\caption{\label{tab:indy1}rates of actions/strategies consistent with reporting true peaks and playing best-response.\\
$a$. Peak reports/dominant strategy play.\\
$b$. Peak reports.\\
$c$. Observed strategy paths consistent moving towards true peak. Value is an upper bound on dominant strategy play due to the possible censoring of the other agent.\\
$d$. Actions consistent with a dominant strategy at a decision node in which the subject is observed.\\
$f.$ \SRUM2  maximizes utility with a $85\%$ frequency, with a range $[76\%,95\%]$ across valuations. Thus, a reasonable measure of rate of empirical best response is the rate of actions that are optimal given \SRUM2 will maximize utility. That is, actions that are part of a perfect equilibrium of the game.
\\
$g$. Frequency of action/path being a best response to the action/path played by the other agent.
}
\end{table}

\begin{table}[t]
\centering
\begin{threeparttable}\footnotesize
\begin{tabular}{lcccc}
&(1)&(2)&(3)&(4)\\
&\begin{tabular}{c}\\reports\\ peak\end{tabular}&\begin{tabular}{c}\\reports\\ peak\end{tabular}&\begin{tabular}{c}plays\\best\\response\end{tabular}&\begin{tabular}{c}plays\\best\\response\end{tabular}\\\hline
second half & 0.057** & 0.056** & 0.036 & 0.021 \\
 & (0.024) & (0.020) & (0.031) & (0.024) \\
\DFUM\ & 0.151** & 0.151** & 0.162*** & 0.162*** \\
 & (0.054) & (0.054) & (0.024) & (0.024) \\
\SRUM1 & -0.006 & -0.006 & -0.108*** & -0.108*** \\
 & (0.043) & (0.043) & (0.028) & (0.028) \\
\SRUM2 & 0.016 & 0.016 & 0.142*** & 0.142*** \\
& (0.046) & (0.046) & (0.017) & (0.017) \\
\OSPUM\ & 0.317*** &  & 0.040 &  \\
node level\tnote{a} & (0.052) &  & (0.033) &  \\
\OSPUM &  & 0.104** &  & -0.094*** \\
path level\tnote{b}   &  & (0.048) &  & (0.026) \\
\hline
observation level&node/action&strategy&node/action&strategy\\
observations& 2,608 & 1,920 & 2,608 & 1,920 \\
r-squared& 0.145 & 0.119 & 0.027 & 0.070\\\hline
\end{tabular}
\begin{tablenotes}
\item \emph{Notes:} \DRUM\ is baseline model. Valuation dummy variables are also included in all regressions. All regression models use cluster-robust standard errors at the session level.
\item[a] Actions consistent with a dominant strategy at a decision node in which the agent is observed.
\item[b] Observed strategy paths consistent with peak reports.
\end{tablenotes}
\end{threeparttable}
\caption{Regression analysis of individual-level decisions by mechanism and role\label{reg:indy}.
}
\end{table}

For all but \SRUM1, reporting one's peak is a dominant strategy; by definition, it also must be a best-response. Independent of mechanism, there may exist other strategies that could still be best responses to certain strategies of the other player. Table \ref{tab:indy1}, panel B provides the rates of all types of best responses (given the actually observed strategy of the other player) for each mechanism and valuation pair. Table \ref{reg:indy}, specifications (3) and (4) provide the corresponding results in regression form. 

The rates of best response are considerably higher than rates of reporting true peaks. Moreover, the differences across treatments change in ordering. Subjects in \DFUM\ and \SRUM2 are both about 15 percentage points more likely to play a best response than in \DRUM\  ($p<0.01$, both comparisons). Interestingly, a subject under \DFUM\ is as likely to play a best response as in \SRUM2 ($p\approx0.414$), a player who observes the other player's move with certainty. Subjects in \OSPUM\ are 9.4 percentage points less likely to choose a best response than under the baseline \DRUM\ mechanisms ($p<0.001$). This rate is statistically indistinguishable from \SRUM1 ($p\approx0.678$). 

\begin{figure}[]
\centering
\begin{pspicture}(0,0)(6,6.5)
\rput[c](3,6.5){\footnotesize (a)}
\rput[c](3,1.2){\footnotesize Report/Award Subject 1}
\rput[c]{90}(0.2,4){\footnotesize Report/Award Subject 2}
\psframe(1,0)(11.2,.8)
\psdots[dotstyle=Bpentagon,dotsize=6pt,linecolor=red](1.7,.4)
\rput[l](2,.4){\footnotesize True Type}
\psdots[dotstyle=square*,linecolor=gray](4,.4)
\rput[l](4.2,.4){\footnotesize True Uniform Outcome}
\psdots[dotsize=2.2pt](8,.4)
\rput[l](8.2,.4){\footnotesize Nash Equilibrium}
\rput(1,1.7){$\mbox{\footnotesize$0$}$}
\rput(2,1.7){$\mbox{\footnotesize$5$}$}
\rput(3,1.7){$\mbox{\footnotesize$10$}$}
\rput(4,1.7){$\mbox{\footnotesize$15$}$}
\rput(5,1.7){$\mbox{\footnotesize$20$}$}
\rput[r](0.8,2){$\mbox{\footnotesize$0$}$}
\rput[r](0.8,3){$\mbox{\footnotesize$5$}$}
\rput[r](0.8,4){$\mbox{\footnotesize$10$}$}
\rput[r](0.8,5){$\mbox{\footnotesize$15$}$}
\rput[r](0.8,6){$\mbox{\footnotesize$20$}$}
\psline[linecolor=red](1,6)(5,2)
\psdots[dotstyle=Bpentagon,dotsize=6pt,linecolor=red](1.4,2.8)
\psdots[dotstyle=square](1,2)
(1.2,2)
(1.4,2)
(1.6,2)
(1.8,2)
(2,2)
(2.2,2)
(2.4,2)
(2.6,2)
(2.8,2)
(3,2)
(3.2,2)
(3.4,2)
(3.6,2)
(3.8,2)
(4,2)
(4.2,2)
(4.4,2)
(4.6,2)
(4.8,2)
(5,2)
(1,2.2)
(1.2,2.2)
(1.4,2.2)
(1.6,2.2)
(1.8,2.2)
(2,2.2)
(2.2,2.2)
(2.4,2.2)
(2.6,2.2)
(2.8,2.2)
(3,2.2)
(3.2,2.2)
(3.4,2.2)
(3.6,2.2)
(3.8,2.2)
(4,2.2)
(4.2,2.2)
(4.4,2.2)
(4.6,2.2)
(4.8,2.2)
(5,2.2)
(1,2.4)
(1.2,2.4)
(1.4,2.4)
(1.6,2.4)
(1.8,2.4)
(2,2.4)
(2.2,2.4)
(2.4,2.4)
(2.6,2.4)
(2.8,2.4)
(3,2.4)
(3.2,2.4)
(3.4,2.4)
(3.6,2.4)
(3.8,2.4)
(4,2.4)
(4.2,2.4)
(4.4,2.4)
(4.6,2.4)
(4.8,2.4)
(5,2.4)
(1,2.6)
(1.2,2.6)
(1.4,2.6)
(1.6,2.6)
(1.8,2.6)
(2,2.6)
(2.2,2.6)
(2.4,2.6)
(2.6,2.6)
(2.8,2.6)
(3,2.6)
(3.2,2.6)
(3.4,2.6)
(3.6,2.6)
(3.8,2.6)
(4,2.6)
(4.2,2.6)
(4.4,2.6)
(4.6,2.6)
(4.8,2.6)
(5,2.6)
(1,2.8)
(1.2,2.8)
(1.4,2.8)
(1.6,2.8)
(1.8,2.8)
(2,2.8)
(2.2,2.8)
(2.4,2.8)
(2.6,2.8)
(2.8,2.8)
(3,2.8)
(3.2,2.8)
(3.4,2.8)
(3.6,2.8)
(3.8,2.8)
(4,2.8)
(4.2,2.8)
(4.4,2.8)
(4.6,2.8)
(4.8,2.8)
(5,2.8)
(1,3)
(1.2,3)
(1.4,3)
(1.6,3)
(1.8,3)
(2,3)
(2.2,3)
(2.4,3)
(2.6,3)
(2.8,3)
(3,3)
(3.2,3)
(3.4,3)
(3.6,3)
(3.8,3)
(4,3)
(4.2,3)
(4.4,3)
(4.6,3)
(4.8,3)
(5,3)
(1,3.2)
(1.2,3.2)
(1.4,3.2)
(1.6,3.2)
(1.8,3.2)
(2,3.2)
(2.2,3.2)
(2.4,3.2)
(2.6,3.2)
(2.8,3.2)
(3,3.2)
(3.2,3.2)
(3.4,3.2)
(3.6,3.2)
(3.8,3.2)
(4,3.2)
(4.2,3.2)
(4.4,3.2)
(4.6,3.2)
(4.8,3.2)
(5,3.2)
(1,3.4)
(1.2,3.4)
(1.4,3.4)
(1.6,3.4)
(1.8,3.4)
(2,3.4)
(2.2,3.4)
(2.4,3.4)
(2.6,3.4)
(2.8,3.4)
(3,3.4)
(3.2,3.4)
(3.4,3.4)
(3.6,3.4)
(3.8,3.4)
(4,3.4)
(4.2,3.4)
(4.4,3.4)
(4.6,3.4)
(4.8,3.4)
(5,3.4)
(1,3.6)
(1.2,3.6)
(1.4,3.6)
(1.6,3.6)
(1.8,3.6)
(2,3.6)
(2.2,3.6)
(2.4,3.6)
(2.6,3.6)
(2.8,3.6)
(3,3.6)
(3.2,3.6)
(3.4,3.6)
(3.6,3.6)
(3.8,3.6)
(4,3.6)
(4.2,3.6)
(4.4,3.6)
(4.6,3.6)
(4.8,3.6)
(5,3.6)
(1,3.8)
(1.2,3.8)
(1.4,3.8)
(1.6,3.8)
(1.8,3.8)
(2,3.8)
(2.2,3.8)
(2.4,3.8)
(2.6,3.8)
(2.8,3.8)
(3,3.8)
(3.2,3.8)
(3.4,3.8)
(3.6,3.8)
(3.8,3.8)
(4,3.8)
(4.2,3.8)
(4.4,3.8)
(4.6,3.8)
(4.8,3.8)
(5,3.8)
(1,4)
(1.2,4)
(1.4,4)
(1.6,4)
(1.8,4)
(2,4)
(2.2,4)
(2.4,4)
(2.6,4)
(2.8,4)
(3,4)
(3.2,4)
(3.4,4)
(3.6,4)
(3.8,4)
(4,4)
(4.2,4)
(4.4,4)
(4.6,4)
(4.8,4)
(5,4)
(1,4.2)
(1.2,4.2)
(1.4,4.2)
(1.6,4.2)
(1.8,4.2)
(2,4.2)
(2.2,4.2)
(2.4,4.2)
(2.6,4.2)
(2.8,4.2)
(3,4.2)
(3.2,4.2)
(3.4,4.2)
(3.6,4.2)
(3.8,4.2)
(4,4.2)
(4.2,4.2)
(4.4,4.2)
(4.6,4.2)
(4.8,4.2)
(5,4.2)
(1,4.4)
(1.2,4.4)
(1.4,4.4)
(1.6,4.4)
(1.8,4.4)
(2,4.4)
(2.2,4.4)
(2.4,4.4)
(2.6,4.4)
(2.8,4.4)
(3,4.4)
(3.2,4.4)
(3.4,4.4)
(3.6,4.4)
(3.8,4.4)
(4,4.4)
(4.2,4.4)
(4.4,4.4)
(4.6,4.4)
(4.8,4.4)
(5,4.4)
(1,4.6)
(1.2,4.6)
(1.4,4.6)
(1.6,4.6)
(1.8,4.6)
(2,4.6)
(2.2,4.6)
(2.4,4.6)
(2.6,4.6)
(2.8,4.6)
(3,4.6)
(3.2,4.6)
(3.4,4.6)
(3.6,4.6)
(3.8,4.6)
(4,4.6)
(4.2,4.6)
(4.4,4.6)
(4.6,4.6)
(4.8,4.6)
(5,4.6)
(1,4.8)
(1.2,4.8)
(1.4,4.8)
(1.6,4.8)
(1.8,4.8)
(2,4.8)
(2.2,4.8)
(2.4,4.8)
(2.6,4.8)
(2.8,4.8)
(3,4.8)
(3.2,4.8)
(3.4,4.8)
(3.6,4.8)
(3.8,4.8)
(4,4.8)
(4.2,4.8)
(4.4,4.8)
(4.6,4.8)
(4.8,4.8)
(5,4.8)
(1,5)
(1.2,5)
(1.4,5)
(1.6,5)
(1.8,5)
(2,5)
(2.2,5)
(2.4,5)
(2.6,5)
(2.8,5)
(3,5)
(3.2,5)
(3.4,5)
(3.6,5)
(3.8,5)
(4,5)
(4.2,5)
(4.4,5)
(4.6,5)
(4.8,5)
(5,5)
(1,5.2)
(1.2,5.2)
(1.4,5.2)
(1.6,5.2)
(1.8,5.2)
(2,5.2)
(2.2,5.2)
(2.4,5.2)
(2.6,5.2)
(2.8,5.2)
(3,5.2)
(3.2,5.2)
(3.4,5.2)
(3.6,5.2)
(3.8,5.2)
(4,5.2)
(4.2,5.2)
(4.4,5.2)
(4.6,5.2)
(4.8,5.2)
(5,5.2)
(1,5.4)
(1.2,5.4)
(1.4,5.4)
(1.6,5.4)
(1.8,5.4)
(2,5.4)
(2.2,5.4)
(2.4,5.4)
(2.6,5.4)
(2.8,5.4)
(3,5.4)
(3.2,5.4)
(3.4,5.4)
(3.6,5.4)
(3.8,5.4)
(4,5.4)
(4.2,5.4)
(4.4,5.4)
(4.6,5.4)
(4.8,5.4)
(5,5.4)
(1,5.6)
(1.2,5.6)
(1.4,5.6)
(1.6,5.6)
(1.8,5.6)
(2,5.6)
(2.2,5.6)
(2.4,5.6)
(2.6,5.6)
(2.8,5.6)
(3,5.6)
(3.2,5.6)
(3.4,5.6)
(3.6,5.6)
(3.8,5.6)
(4,5.6)
(4.2,5.6)
(4.4,5.6)
(4.6,5.6)
(4.8,5.6)
(5,5.6)
(1,5.8)
(1.2,5.8)
(1.4,5.8)
(1.6,5.8)
(1.8,5.8)
(2,5.8)
(2.2,5.8)
(2.4,5.8)
(2.6,5.8)
(2.8,5.8)
(3,5.8)
(3.2,5.8)
(3.4,5.8)
(3.6,5.8)
(3.8,5.8)
(4,5.8)
(4.2,5.8)
(4.4,5.8)
(4.6,5.8)
(4.8,5.8)
(5,5.8)
(1,6)
(1.2,6)
(1.4,6)
(1.6,6)
(1.8,6)
(2,6)
(2.2,6)
(2.4,6)
(2.6,6)
(2.8,6)
(3,6)
(3.2,6)
(3.4,6)
(3.6,6)
(3.8,6)
(4,6)
(4.2,6)
(4.4,6)
(4.6,6)
(4.8,6)
(5,6)
\psdots[dotstyle=square*,linecolor=gray](1,2)
(1.2,2)
(1.4,2)
(1.6,2)
(1.8,2)
(2,2)
(2.2,2)
(2.4,2)
(2.6,2)
(2.8,2)
(3,2)
(1,2.2)
(1.2,2.2)
(1.4,2.2)
(1.6,2.2)
(1.8,2.2)
(2,2.2)
(2.2,2.2)
(2.4,2.2)
(2.6,2.2)
(2.8,2.2)
(3,2.2)
(1,2.4)
(1.2,2.4)
(1.4,2.4)
(1.6,2.4)
(1.8,2.4)
(2,2.4)
(2.2,2.4)
(2.4,2.4)
(2.6,2.4)
(2.8,2.4)
(3,2.4)
(1,2.6)
(1.2,2.6)
(1.4,2.6)
(1.6,2.6)
(1.8,2.6)
(2,2.6)
(2.2,2.6)
(2.4,2.6)
(2.6,2.6)
(2.8,2.6)
(3,2.6)
(1,2.8)
(1.2,2.8)
(1.4,2.8)
(1.6,2.8)
(1.8,2.8)
(2,2.8)
(2.2,2.8)
(2.4,2.8)
(2.6,2.8)
(2.8,2.8)
(3,2.8)
(1,3)
(1.2,3)
(1.4,3)
(1.6,3)
(1.8,3)
(2,3)
(2.2,3)
(2.4,3)
(2.6,3)
(2.8,3)
(3,3)
(1,3.2)
(1.2,3.2)
(1.4,3.2)
(1.6,3.2)
(1.8,3.2)
(2,3.2)
(2.2,3.2)
(2.4,3.2)
(2.6,3.2)
(2.8,3.2)
(3,3.2)
(1,3.4)
(1.2,3.4)
(1.4,3.4)
(1.6,3.4)
(1.8,3.4)
(2,3.4)
(2.2,3.4)
(2.4,3.4)
(2.6,3.4)
(2.8,3.4)
(3,3.4)
(1,3.6)
(1.2,3.6)
(1.4,3.6)
(1.6,3.6)
(1.8,3.6)
(2,3.6)
(2.2,3.6)
(2.4,3.6)
(2.6,3.6)
(2.8,3.6)
(3,3.6)
(1,3.8)
(1.2,3.8)
(1.4,3.8)
(1.6,3.8)
(1.8,3.8)
(2,3.8)
(2.2,3.8)
(2.4,3.8)
(2.6,3.8)
(2.8,3.8)
(3,3.8)
(1,4)
(1.2,4)
(1.4,4)
(1.6,4)
(1.8,4)
(2,4)
(2.2,4)
(2.4,4)
(2.6,4)
(2.8,4)
(3,4)
(3,4)
(3.2,4)
(3.4,4)
(3.6,4)
(3.8,4)
(4,4)
(4.2,4)
(4.4,4)
(4.6,4)
(4.8,4)
(5,4)
(3,4.2)
(3.2,4.2)
(3.4,4.2)
(3.6,4.2)
(3.8,4.2)
(4,4.2)
(4.2,4.2)
(4.4,4.2)
(4.6,4.2)
(4.8,4.2)
(5,4.2)
(3,4.4)
(3.2,4.4)
(3.4,4.4)
(3.6,4.4)
(3.8,4.4)
(4,4.4)
(4.2,4.4)
(4.4,4.4)
(4.6,4.4)
(4.8,4.4)
(5,4.4)
(3,4.6)
(3.2,4.6)
(3.4,4.6)
(3.6,4.6)
(3.8,4.6)
(4,4.6)
(4.2,4.6)
(4.4,4.6)
(4.6,4.6)
(4.8,4.6)
(5,4.6)
(3,4.8)
(3.2,4.8)
(3.4,4.8)
(3.6,4.8)
(3.8,4.8)
(4,4.8)
(4.2,4.8)
(4.4,4.8)
(4.6,4.8)
(4.8,4.8)
(5,4.8)
(3,5)
(3.2,5)
(3.4,5)
(3.6,5)
(3.8,5)
(4,5)
(4.2,5)
(4.4,5)
(4.6,5)
(4.8,5)
(5,5)
(3,5.2)
(3.2,5.2)
(3.4,5.2)
(3.6,5.2)
(3.8,5.2)
(4,5.2)
(4.2,5.2)
(4.4,5.2)
(4.6,5.2)
(4.8,5.2)
(5,5.2)
(3,5.4)
(3.2,5.4)
(3.4,5.4)
(3.6,5.4)
(3.8,5.4)
(4,5.4)
(4.2,5.4)
(4.4,5.4)
(4.6,5.4)
(4.8,5.4)
(5,5.4)
(3,5.6)
(3.2,5.6)
(3.4,5.6)
(3.6,5.6)
(3.8,5.6)
(4,5.6)
(4.2,5.6)
(4.4,5.6)
(4.6,5.6)
(4.8,5.6)
(5,5.6)
(3,5.8)
(3.2,5.8)
(3.4,5.8)
(3.6,5.8)
(3.8,5.8)
(4,5.8)
(4.2,5.8)
(4.4,5.8)
(4.6,5.8)
(4.8,5.8)
(5,5.8)
(3,6)
(3.2,6)
(3.4,6)
(3.6,6)
(3.8,6)
(4,6)
(4.2,6)
(4.4,6)
(4.6,6)
(4.8,6)
(5,6)

\psdots[dotsize=2.2pt](1,2)
(1.2,2)
(1.4,2)
(1.6,2)
(1.8,2)
(2,2)
(2.2,2)
(2.4,2)
(2.6,2)
(2.8,2)
(3,2)
(1,2.2)
(1.2,2.2)
(1.4,2.2)
(1.6,2.2)
(1.8,2.2)
(2,2.2)
(2.2,2.2)
(2.4,2.2)
(2.6,2.2)
(2.8,2.2)
(3,2.2)
(1,2.4)
(1.2,2.4)
(1.4,2.4)
(1.6,2.4)
(1.8,2.4)
(2,2.4)
(2.2,2.4)
(2.4,2.4)
(2.6,2.4)
(2.8,2.4)
(3,2.4)
(1,2.6)
(1.2,2.6)
(1.4,2.6)
(1.6,2.6)
(1.8,2.6)
(2,2.6)
(2.2,2.6)
(2.4,2.6)
(2.6,2.6)
(2.8,2.6)
(3,2.6)
(1,2.8)
(1.2,2.8)
(1.4,2.8)
(1.6,2.8)
(1.8,2.8)
(2,2.8)
(2.2,2.8)
(2.4,2.8)
(2.6,2.8)
(2.8,2.8)
(3,2.8)
(1,3)
(1.2,3)
(1.4,3)
(1.6,3)
(1.8,3)
(2,3)
(2.2,3)
(2.4,3)
(2.6,3)
(2.8,3)
(3,3)
(1,3.2)
(1.2,3.2)
(1.4,3.2)
(1.6,3.2)
(1.8,3.2)
(2,3.2)
(2.2,3.2)
(2.4,3.2)
(2.6,3.2)
(2.8,3.2)
(3,3.2)
(1,3.4)
(1.2,3.4)
(1.4,3.4)
(1.6,3.4)
(1.8,3.4)
(2,3.4)
(2.2,3.4)
(2.4,3.4)
(2.6,3.4)
(2.8,3.4)
(3,3.4)
(1,3.6)
(1.2,3.6)
(1.4,3.6)
(1.6,3.6)
(1.8,3.6)
(2,3.6)
(2.2,3.6)
(2.4,3.6)
(2.6,3.6)
(2.8,3.6)
(3,3.6)
(1,3.8)
(1.2,3.8)
(1.4,3.8)
(1.6,3.8)
(1.8,3.8)
(2,3.8)
(2.2,3.8)
(2.4,3.8)
(2.6,3.8)
(2.8,3.8)
(3,3.8)
(1,4)
(1.2,4)
(1.4,4)
(1.6,4)
(1.8,4)
(2,4)
(2.2,4)
(2.4,4)
(2.6,4)
(2.8,4)
(3,4)
\end{pspicture}
\begin{pspicture}(0,0)(5,6)
\rput[c](3,6.5){\footnotesize (b)}
\rput[c](3,1.2){\footnotesize Report/Award Subject 1}
\rput[c]{90}(0.2,4){\footnotesize Report/Award Subject 2}
\rput(1,1.7){$\mbox{\footnotesize$0$}$}
\rput(2,1.7){$\mbox{\footnotesize$5$}$}
\rput(3,1.7){$\mbox{\footnotesize$10$}$}
\rput(4,1.7){$\mbox{\footnotesize$15$}$}
\rput(5,1.7){$\mbox{\footnotesize$20$}$}
\rput[r](0.8,2){$\mbox{\footnotesize$0$}$}
\rput[r](0.8,3){$\mbox{\footnotesize$5$}$}
\rput[r](0.8,4){$\mbox{\footnotesize$10$}$}
\rput[r](0.8,5){$\mbox{\footnotesize$15$}$}
\rput[r](0.8,6){$\mbox{\footnotesize$20$}$}
\psline[linecolor=red](1,6)(5,2)
\psdots[dotstyle=Bpentagon,dotsize=6pt,linecolor=red](3.6,2.6)
\psdots[dotstyle=square](1,2)
(1.2,2)
(1.4,2)
(1.6,2)
(1.8,2)
(2,2)
(2.2,2)
(2.4,2)
(2.6,2)
(2.8,2)
(3,2)
(3.2,2)
(3.4,2)
(3.6,2)
(3.8,2)
(4,2)
(4.2,2)
(4.4,2)
(4.6,2)
(4.8,2)
(5,2)
(1,2.2)
(1.2,2.2)
(1.4,2.2)
(1.6,2.2)
(1.8,2.2)
(2,2.2)
(2.2,2.2)
(2.4,2.2)
(2.6,2.2)
(2.8,2.2)
(3,2.2)
(3.2,2.2)
(3.4,2.2)
(3.6,2.2)
(3.8,2.2)
(4,2.2)
(4.2,2.2)
(4.4,2.2)
(4.6,2.2)
(4.8,2.2)
(5,2.2)
(1,2.4)
(1.2,2.4)
(1.4,2.4)
(1.6,2.4)
(1.8,2.4)
(2,2.4)
(2.2,2.4)
(2.4,2.4)
(2.6,2.4)
(2.8,2.4)
(3,2.4)
(3.2,2.4)
(3.4,2.4)
(3.6,2.4)
(3.8,2.4)
(4,2.4)
(4.2,2.4)
(4.4,2.4)
(4.6,2.4)
(4.8,2.4)
(5,2.4)
(1,2.6)
(1.2,2.6)
(1.4,2.6)
(1.6,2.6)
(1.8,2.6)
(2,2.6)
(2.2,2.6)
(2.4,2.6)
(2.6,2.6)
(2.8,2.6)
(3,2.6)
(3.2,2.6)
(3.4,2.6)
(3.6,2.6)
(3.8,2.6)
(4,2.6)
(4.2,2.6)
(4.4,2.6)
(4.6,2.6)
(4.8,2.6)
(5,2.6)
(1,2.8)
(1.2,2.8)
(1.4,2.8)
(1.6,2.8)
(1.8,2.8)
(2,2.8)
(2.2,2.8)
(2.4,2.8)
(2.6,2.8)
(2.8,2.8)
(3,2.8)
(3.2,2.8)
(3.4,2.8)
(3.6,2.8)
(3.8,2.8)
(4,2.8)
(4.2,2.8)
(4.4,2.8)
(4.6,2.8)
(4.8,2.8)
(5,2.8)
(1,3)
(1.2,3)
(1.4,3)
(1.6,3)
(1.8,3)
(2,3)
(2.2,3)
(2.4,3)
(2.6,3)
(2.8,3)
(3,3)
(3.2,3)
(3.4,3)
(3.6,3)
(3.8,3)
(4,3)
(4.2,3)
(4.4,3)
(4.6,3)
(4.8,3)
(5,3)
(1,3.2)
(1.2,3.2)
(1.4,3.2)
(1.6,3.2)
(1.8,3.2)
(2,3.2)
(2.2,3.2)
(2.4,3.2)
(2.6,3.2)
(2.8,3.2)
(3,3.2)
(3.2,3.2)
(3.4,3.2)
(3.6,3.2)
(3.8,3.2)
(4,3.2)
(4.2,3.2)
(4.4,3.2)
(4.6,3.2)
(4.8,3.2)
(5,3.2)
(1,3.4)
(1.2,3.4)
(1.4,3.4)
(1.6,3.4)
(1.8,3.4)
(2,3.4)
(2.2,3.4)
(2.4,3.4)
(2.6,3.4)
(2.8,3.4)
(3,3.4)
(3.2,3.4)
(3.4,3.4)
(3.6,3.4)
(3.8,3.4)
(4,3.4)
(4.2,3.4)
(4.4,3.4)
(4.6,3.4)
(4.8,3.4)
(5,3.4)
(1,3.6)
(1.2,3.6)
(1.4,3.6)
(1.6,3.6)
(1.8,3.6)
(2,3.6)
(2.2,3.6)
(2.4,3.6)
(2.6,3.6)
(2.8,3.6)
(3,3.6)
(3.2,3.6)
(3.4,3.6)
(3.6,3.6)
(3.8,3.6)
(4,3.6)
(4.2,3.6)
(4.4,3.6)
(4.6,3.6)
(4.8,3.6)
(5,3.6)
(1,3.8)
(1.2,3.8)
(1.4,3.8)
(1.6,3.8)
(1.8,3.8)
(2,3.8)
(2.2,3.8)
(2.4,3.8)
(2.6,3.8)
(2.8,3.8)
(3,3.8)
(3.2,3.8)
(3.4,3.8)
(3.6,3.8)
(3.8,3.8)
(4,3.8)
(4.2,3.8)
(4.4,3.8)
(4.6,3.8)
(4.8,3.8)
(5,3.8)
(1,4)
(1.2,4)
(1.4,4)
(1.6,4)
(1.8,4)
(2,4)
(2.2,4)
(2.4,4)
(2.6,4)
(2.8,4)
(3,4)
(3.2,4)
(3.4,4)
(3.6,4)
(3.8,4)
(4,4)
(4.2,4)
(4.4,4)
(4.6,4)
(4.8,4)
(5,4)
(1,4.2)
(1.2,4.2)
(1.4,4.2)
(1.6,4.2)
(1.8,4.2)
(2,4.2)
(2.2,4.2)
(2.4,4.2)
(2.6,4.2)
(2.8,4.2)
(3,4.2)
(3.2,4.2)
(3.4,4.2)
(3.6,4.2)
(3.8,4.2)
(4,4.2)
(4.2,4.2)
(4.4,4.2)
(4.6,4.2)
(4.8,4.2)
(5,4.2)
(1,4.4)
(1.2,4.4)
(1.4,4.4)
(1.6,4.4)
(1.8,4.4)
(2,4.4)
(2.2,4.4)
(2.4,4.4)
(2.6,4.4)
(2.8,4.4)
(3,4.4)
(3.2,4.4)
(3.4,4.4)
(3.6,4.4)
(3.8,4.4)
(4,4.4)
(4.2,4.4)
(4.4,4.4)
(4.6,4.4)
(4.8,4.4)
(5,4.4)
(1,4.6)
(1.2,4.6)
(1.4,4.6)
(1.6,4.6)
(1.8,4.6)
(2,4.6)
(2.2,4.6)
(2.4,4.6)
(2.6,4.6)
(2.8,4.6)
(3,4.6)
(3.2,4.6)
(3.4,4.6)
(3.6,4.6)
(3.8,4.6)
(4,4.6)
(4.2,4.6)
(4.4,4.6)
(4.6,4.6)
(4.8,4.6)
(5,4.6)
(1,4.8)
(1.2,4.8)
(1.4,4.8)
(1.6,4.8)
(1.8,4.8)
(2,4.8)
(2.2,4.8)
(2.4,4.8)
(2.6,4.8)
(2.8,4.8)
(3,4.8)
(3.2,4.8)
(3.4,4.8)
(3.6,4.8)
(3.8,4.8)
(4,4.8)
(4.2,4.8)
(4.4,4.8)
(4.6,4.8)
(4.8,4.8)
(5,4.8)
(1,5)
(1.2,5)
(1.4,5)
(1.6,5)
(1.8,5)
(2,5)
(2.2,5)
(2.4,5)
(2.6,5)
(2.8,5)
(3,5)
(3.2,5)
(3.4,5)
(3.6,5)
(3.8,5)
(4,5)
(4.2,5)
(4.4,5)
(4.6,5)
(4.8,5)
(5,5)
(1,5.2)
(1.2,5.2)
(1.4,5.2)
(1.6,5.2)
(1.8,5.2)
(2,5.2)
(2.2,5.2)
(2.4,5.2)
(2.6,5.2)
(2.8,5.2)
(3,5.2)
(3.2,5.2)
(3.4,5.2)
(3.6,5.2)
(3.8,5.2)
(4,5.2)
(4.2,5.2)
(4.4,5.2)
(4.6,5.2)
(4.8,5.2)
(5,5.2)
(1,5.4)
(1.2,5.4)
(1.4,5.4)
(1.6,5.4)
(1.8,5.4)
(2,5.4)
(2.2,5.4)
(2.4,5.4)
(2.6,5.4)
(2.8,5.4)
(3,5.4)
(3.2,5.4)
(3.4,5.4)
(3.6,5.4)
(3.8,5.4)
(4,5.4)
(4.2,5.4)
(4.4,5.4)
(4.6,5.4)
(4.8,5.4)
(5,5.4)
(1,5.6)
(1.2,5.6)
(1.4,5.6)
(1.6,5.6)
(1.8,5.6)
(2,5.6)
(2.2,5.6)
(2.4,5.6)
(2.6,5.6)
(2.8,5.6)
(3,5.6)
(3.2,5.6)
(3.4,5.6)
(3.6,5.6)
(3.8,5.6)
(4,5.6)
(4.2,5.6)
(4.4,5.6)
(4.6,5.6)
(4.8,5.6)
(5,5.6)
(1,5.8)
(1.2,5.8)
(1.4,5.8)
(1.6,5.8)
(1.8,5.8)
(2,5.8)
(2.2,5.8)
(2.4,5.8)
(2.6,5.8)
(2.8,5.8)
(3,5.8)
(3.2,5.8)
(3.4,5.8)
(3.6,5.8)
(3.8,5.8)
(4,5.8)
(4.2,5.8)
(4.4,5.8)
(4.6,5.8)
(4.8,5.8)
(5,5.8)
(1,6)
(1.2,6)
(1.4,6)
(1.6,6)
(1.8,6)
(2,6)
(2.2,6)
(2.4,6)
(2.6,6)
(2.8,6)
(3,6)
(3.2,6)
(3.4,6)
(3.6,6)
(3.8,6)
(4,6)
(4.2,6)
(4.4,6)
(4.6,6)
(4.8,6)
(5,6)
\psdots[dotstyle=square*,linecolor=gray](3.6,2)
(3.6,2.2)(3.6,2.4)(3.6,2.6)(3.6,2.8)(3.6,3)(3.6,3.2)(3.6,3.4)
(3.8,3.4)(4,3.4)(4.2,3.4)(4.4,3.4)(4.6,3.4)(4.8,3.4)(5,3.4)
\psdots[dotsize=2.2pt](3.6,2)
(3.6,2.2)(3.6,2.4)(3.6,2.6)(3.6,2.8)(3.6,3)(3.6,3.2)(3.6,3.4)
(3.4,3.6)
(3.2,3.8)(3,4)
(1,4)(1.2,4)(1.4,4)(1.6,4)(1.8,4)(2,4)
(2.2,4)(2.4,4)(2.6,4)(2.8,4)
(3,4.2)(3,4.4)(3,4.6)(3,4.8)(3,5)(3,5.2)(3,5.4)(3,5.6)(3,5.8)(3,6)
\end{pspicture}
\caption{Uniform outcome and Nash regions for the Uniform rule. (a, left) Peak profile $(3,4)$ (valuation 1). Uniform outcome is produced by 54.65\% of possible reports; 27.44\% of reports are Nash equilibria. All Nash equilibria produce the Uniform outcome. (b, right) Peak profile $(13,3)$ (valuation 4). Uniform outcome is produced by 3.4\% of reports; 8.6\% of profiles are Nash equilibria, of which 39\% produce the Uniform outcome.}
\label{Fig:Uniform2}
\end{figure}
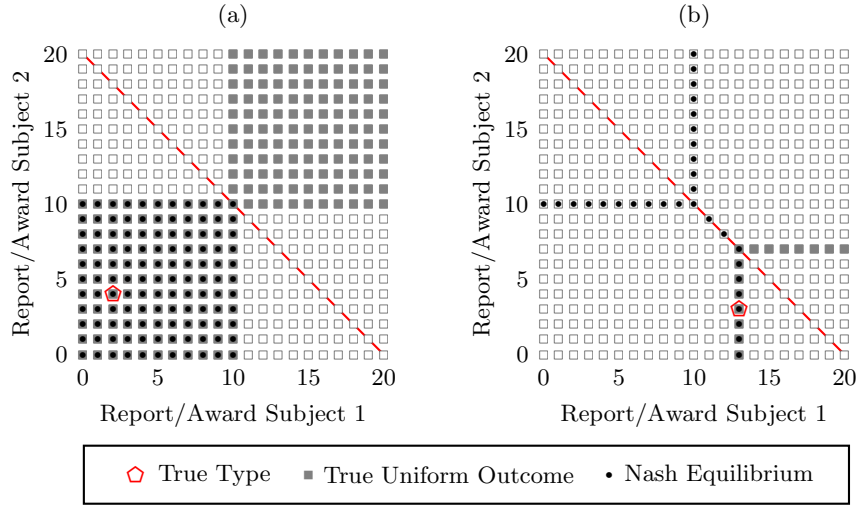

The overall rate of best response in \DRUM\ squared ($0.713^2=0.508$) is roughly equal to the rate of Uniform outcomes achieved overall in that mechanism (0.487, see Table \ref{overall}), but both measures appear to vary greatly across valuations and are seemingly unrelated with the frequency of peak reports. By design, the single profile of true peaks is always both a Nash equilibrium profile and produces the Uniform outcomes. However, these latter two sets of profiles generally do not coincide. 
Figure~\ref{Fig:Uniform2} illustrates the shape and size of these sets for valuations $(3,4)$ and $(13,3)$.

Consider first a profile of true peaks $(3,4)$ for which $U$ recommends the equal division, $(10,10)$ (see Figure~\ref{Fig:Uniform2}(a)). The set of reports that produce this outcome is composed by two quadrants of the set of possible reports, a total of 54.65\%. From this set, about half  constitute Nash equilibria of \DRUM. By contrast, when $U$ does not recommend equal division, these sets are small. For instance, when true peaks are $(13,3)$ as in Figure~\ref{Fig:Uniform2}(b), the set of profiles that produce $U$ is $L$-shaped: either the first agent reports $13$ and the second between $0$ and $7$, or the first agent reports between $14$ and $20$ and the second agent reports $7$. The set of Nash equilibria that produce $U$ in this case is the segment in which the first agent reports $13$ and the second between $0$ and $7$. There are 21 additional Nash equilibrium profiles.

Table~\ref{tab:PredictedUDRU} presents the proportion of profiles that produce $U$ across valuations. For the first two valuations a majority of profiles produces the Uniform outcome. For valuations $3$--$6$, only a narrow set of profiles does it. If one accounts for these differences, the apparent erratic relation between peak-reporting and mechanism performance is reconciled to a great extent. One can expect a high frequency of Uniform outcomes in \DRUM\ when the intended outcome is equal division; moreover, the achievement of nearly all of these Uniform outcomes does not require peak reports.

We now see that in general, profiles that produce Uniform outcomes cover a much greater percentage of strategy space than profiles that are Nash equilibria, but this difference is almost entirely due to valuations 1 and 2. In valuations 3--6, the set of Nash profiles is larger. In contrast with these two larger sets, the single profile that is the set of truthful reports only occupies $1/441$ of the strategy space. The frequency of subject play for all three categories of profiles in higher than random for all but one treatment and valuation combination.\footnote{No peak report profiles were obtained for the \OSPUM\ in valuation 3. (See table \ref{tab:Aprofiles}).}

\begin{table}[]
\centering
\begin{threeparttable}\footnotesize
\begin{tabular}{@{}p{0.95\textwidth}@{}}
\centering
\begin{tabular}{ccccccc}
\toprule
 &\multirow{4}{*}{\begin{tabular}{c}share of\\ profiles\\ that\\ produce $U$\end{tabular}}   & \multirow{4}{*}{\begin{tabular}{c}share of\\ profiles\\ that\\ are Nash\end{tabular}}& \multirow{4}{*}{\begin{tabular}{c}share of\\ profiles with \\ both peak \\ reports\end{tabular}}&\multicolumn{3}{c}{\begin{tabular}{c}frequency of observation\\ (\DRUM\ only)\end{tabular}}\\
 \cmidrule{5-7}
 &  & &  &U& Nash &peak \\
valuation &  &  &&outcome & profile  &reports\\
\midrule
1 & 0.546 & 0.274 & 0.002& 0.788 & 0.673 &0.077 \\
2 & 0.546 & 0.274 &0.002& 0.865 & 0.808 &0.096 \\
3 & 0.020  & 0.079  &0.002& 0.269  & 0.288  & 0.231 \\
4 & 0.034  & 0.086  &0.002& 0.423  & 0.442  & 0.135 \\
5 & 0.025  & 0.082  &0.002& 0.212  & 0.231 & 0.096  \\
6 & 0.043  & 0.091  &0.002& 0.365  & 0.404 & 0.212\\
\midrule
overall & 0.203  & 0.118  & 0.002 & 0.487  & 0.474 & 0.141\\
\bottomrule
\end{tabular}
%
\end{tabular}
\end{threeparttable}
\caption{Share of profile space (i) that produces Uniform outcome, (ii) that is Nash, and (iii) that is the peak reporting profile and frequency of the observed profiles. Only \DRUM\ treatment shown. See Table \ref{tab:Aprofiles} for other treatments. \label{tab:PredictedUDRU}}
\end{table}

\subsection{The success of \DFUM }

\begin{result}\label{res:DFUM} For the \DFUM : 
\begin{enumerate}[label=\alph*.]
\item Non-binding tentative reports during the reporting period of \DFUM\ largely resemble final reports.
\item Subjects frequently best respond to the non-binding reports they observe, leading to high rates of best response in \DFUM\ relative to \DRUM\ and \SRUM1.  
\item Tentative reports vary over the reporting period and peak reports are best-responses to every possible partner report, so the peak-reporting rate is higher in \DFUM\ than for \SRUM2 .
\end{enumerate}
\end{result}

\begin{figure}
\centering
\includegraphics[width=0.7\textwidth]{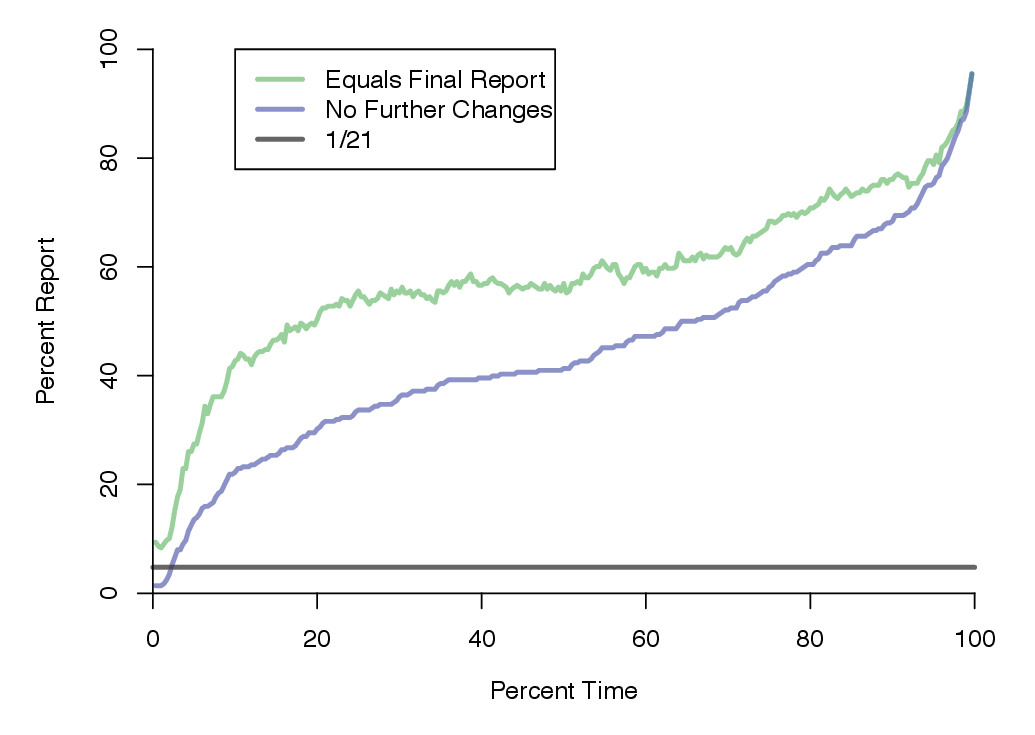}
\caption{Percentage tentative reports that match their finalized report and the percentage of tentative reports that remain unchanged and become finalized reports. The horizontal line represents the equivalent value of random play (1/21).\label{fig:lastone}}
\end{figure}

In all treatments, subjects could freely adjust their report throughout their reporting period, but the information provided to subjects during the reporting period varied between treatments. 
Unlike the other treatments, players in \DFUM\ (\SRUM2) could see the other player's tentative (finalized) report and their own potential payoffs under that report (Figure \ref{fig:ScreenshotCFU}). 
In \DFUM , tentative reports are highly correlated with final reports. Figure \ref{fig:lastone} shows the percentage tentative reports that match finalized reports and the percentage of tentative reports that no longer change for each second of the reporting period. By both measures, tentative reports become exceedingly predictive of final play over time.

Sufficiently sophisticated agents who fully understand the Uniform rule should be able to identify weakly dominant strategies and equilibria once they know their peak and their partner's peak. However, less sophisticated agents who lack ``the ability or inclination to go through any complex reasoning process'' may instead rely on accumulating ``empirical information on the relative advantages of the various pure strategies at their disposal'' \citep{nash1950dissertation}. Subjects in \DFUM\ and \SRUM2 need not  form conjectures about their opponent's report because they can directly observe empirical information about their opponent's report. This empirical information may have made it easier for less sophisticated agents to identify their best responses in \DFUM\ and \SRUM 2. Consistent with this hypothesis, subjects were significantly more likely to select best responses in \DFUM\ and \SRUM 2 than \DRUM\ or \SRUM 1 (see Result \ref{Res:individual}).  

As illustrated in Figure \ref{fig:ScreenshotCFU}, agents can have multiple best responses under the Uniform rule. An agent who knows the strategies employed by others with certainty has no incentive to select a weakly dominant strategy over any other best response. Since \SRUM2 subjects could directly observe the other player's finalized report, they may have seen little advantage in selecting their peak over any other payoff maximizing report. Finalized \SRUM1 reports remained constant over the \SRUM2 reporting period, but the tentative reports observed by subjects in \DFUM\ could be freely adjusted over the reporting period. The peak report is a best response to every possible partner report, so the variability of tentative partner reports over the reporting period in \DFUM\ provided empirical evidence in favor of peak reporting which may have encouraged subjects to select their peak even if they were unable to recognize weak dominance through counterfactual reasoning. Consistent with this hypothesis, subjects in \DFUM\ were significantly more likely to select their peak report than in \SRUM2 (see Result \ref{Res:individual}). 

To formalize this intuition, we introduce the notion of an ``empirically optimal'' report. A given report is said to be empirically optimal for a given subject in a given reporting period if it maximizes their payoff against every partner report they observe over the entire reporting period. In \DFUM\ and SRU2, the peak report is always empirically optimal because it is a dominant strategy, but other reporting strategies may also be empirically optimal. Since observed partner reports could vary over the reporting period in \DFUM\ but not in \SRUM 2, a larger percentage of empirically optimal reports were peak reports in the former. In 73\% of periods in \DFUM, the only empirically optimal report was the dominant strategy peak report. Conversely, this occurred in only 13\% of periods in \SRUM 2. This large discrepancy may explain why \DFUM\ subjects were more likely to select peak reports.

To estimate the relative influence of dominance and empirical optimality on subject behavior, we estimate a structural model. The model aims to predict a subject's final report from the partner reports they observe observes during their reporting period. In \DFUM\ these are tentative partner reports, while in \SRUM 2 these are finalized, first-mover reports. Formally, let $\theta_{it}\in\Theta$ denote agent $i$'s preferred amount in period $t$. Let $D_{it}:\Theta\to\{0,1\}$ indicate weak dominance with $D_{it}(x)=1$ if $x=\theta_{it}$ and $D_{it}(x)=0$ otherwise. Similarly, let $E_{it}:\Theta\to\{0,1\}$ indicate empirical optimality with $E_{it}(x)=1$ if $x$ is empirically optimal for subject $i$ in period $t$ and $E_{it}(x)=0$ otherwise. Let $A_{it}(x)$ denote subject $i$'s attraction to report $x\in\Theta$ in period $t$ such that $A_{it}(x)=\lambda_{E}E_{it}(x)+\lambda_{D}D_{it}(x)$. Here $\lambda_E$ denotes sensitivity to empirical optimality and $\lambda_D$ denotes sensitivity to weak dominance. Let $P_{it}(x)$ denote the probability that subject $i$ selects finalized report $x\in\Theta$ in period $t$, given by:
\begin{equation}
P_{it}(x)=\frac{\exp A_{i}\left(x\right)}{\sum_{y=0}^{20}\exp A_{i}\left(y\right)}
\end{equation}

\begin{table}
\centering
\begin{tabular}{rcccccc}
&\multicolumn{2}{c}{overall}&\multicolumn{2}{c}{\DFUM\ }&\multicolumn{2}{c}{\SRUM2} \\
          & $\lambda_E$ & $\lambda_D$ & $\lambda_E$ & $\lambda_D$ & $\lambda_E$ & $\lambda_D$\\
\hline
Estimate  &  1.41       &  1.79&0.734&2.352&1.850&1.656       \\
Std. Err. & (0.31)      & (0.20)& (0.176)&(0.091)&(0.136)&(0.116)     \\
\hline
\end{tabular}
\caption{Maximum likelihood estimates and session-level bootstrap standard errors\label{tab:MLE}}
\end{table}

Table \ref{tab:MLE} presents maximum likelihood estimates for this model. The leftmost columns present parameter estimates for the combination \DFUM\ and \SRUM 2. The middle columns present estimates for \DFUM\ specifically. The rightmost columns present estimates for \SRUM2 specifically. All specifications suggest that both empirical optimality and weak dominance had significant effects on finalized reports. Under the main specification, this model predicts 36.5\% peak reporting for \SRUM2 and 50.6\% peak reporting in \DFUM , broadly in line with the observed peak reporting rates. This model predicts that subjects in \DFUM\ are more likely to select peak reports than in \SRUM2 because peak reports form a larger share of empirically optimal reports.

The model estimates are also shown restricted to each specific treatment. We see that in both treatments, subjects are attracted to the dominant-strategy peak report independent of its being empirically optimal (i.e., $\lambda_D>0$). This effect is larger for \DFUM\ than \SRUM2. The effect relative to $\lambda_E$ is even greater in \DFUM\ treatment. The dominant strategy may have been more appealing in \DFUM\ than for \SRUM2 because strategic uncertainty is present in \DFUM\ but not in \SRUM2. In \DFUM\ a subject can directly observe the strategic uncertainty they encounter. A peak report is particularly useful in that it never needs to be changed in response to changes in tentative partner reports.

Thus, the success of \DFUM\ is the combination of two effects. A high frequency of best responses due to its informational simplicity; and a higher frequency of peak reports, possibly due to the direct evidence it provides about the superiority of peak reports. Since $U$ is non-bossy, a profile of in which one agent best responds to the other agent's truthful report necessarily produces the Uniform outcome for true types 
\citep{Schummer-Velez-2019,Bochet-Tummenassan-2020}. Thus, gains in truthful reporting translate into significant performance gains.

\subsection{The challenges of \OSPUM}

\begin{result}\label{res:OSPUM} For the \OSPUM\ :
\begin{enumerate}[label=\alph*.]
\item Subjects struggle most to play actions consistent with dominant strategies at the initial node and when doing so requires stopping the mechanism. Together, these factors lead to several cases of the mechanism moving in the direction opposite the equilibrium path after the initial node.
\item For nodes where subjects are still moving towards their peak, subjects have very high rates of non-termination.
\item Subjects are more likely to play the action associated with a dominant strategy when part of plan of action that is 0-step rather than 1-step Simply Dominant. 
\end{enumerate}
\end{result}

\DRUM, \SRUM, and \DFUM\ differ from \OSPUM\ in that truthful reporting of one's peak requires a single action at a single node. 
In \OSPUM\ a peak report corresponds to a choice of multiple actions over multiple nodes towards the player's peak. 
While the dominant-strategy equilibrium path in valuations 1 and 2 requires play at only one node, valuation 3--6 require play at 7, 4, 6 and 2, nodes respectively. 
A quick look at Table \ref{overall} shows how this property alters outcomes. \OSPUM\ achieves Uniform outcomes at a rate of 0.717 for the first two valuations and 0.300 for the latter four ($p<0.1$, two-tailed binomial test at the session level). 

The higher rate of Uniform outcomes in valuations 1 and 2 is not due to a particularly high rate of equilibrium actions at the initial node. In the 92 observed cases of the Uniform outcome in valuations 1 and 2, only 21 (20.6\%) of those cases occurred because both players played an action consistent with reporting their peaks.\footnote{In the other 72, in 52 (56.5\%) of those cases only one player played an action consistent with peak reporting and in the other 21 neither player played such an action, but their actions forced the mechanism to terminate, nonetheless.} At the initial node, subjects play actions consistent with peak reporting at relatively low rates (312 of 552 cases, 56.5\%).

Table \ref{tab:sumOSP} provides a characterization of all node-level observations under the \OSPUM\ mechanism. Given 46 subjects, 2 repetitions of valuations and 21 expected node observations over the six valuations ($1+1+7+4+6+2$, see above), we should have observed 552 observations at the initial node and 1380 ($46\times 2 \times (21-6)$) observations at later nodes. Instead of 1,932, we only observed 1,240 due to play that deviated from the equilibrium path. Actions inconsistent with playing a dominant strategy are most common at initial nodes (as aforementioned) and in cases where the player should terminate the mechanism (i.e., at one's peak, past one's peak, and moving away from one's peak after the initial node, 41.4\% combined rate).
\begin{table}[]
\centering
\resizebox{1\textwidth}{!} {
\begin{tabular}{lccc}
\toprule
node classification & \begin{tabular}{c}number of\\observations \end{tabular} & \begin{tabular}{c}plays dominant-\\strategy action\end{tabular}& percentage  \\
\midrule
initial node, 1-step Simply Dominant only & 138 (\emph{138})              &84  (\emph{138})              &60.9\%  (\emph{100\%})               \\
initial node, 0-step Simply Dominant & 414 (\emph{414})              &228 (\emph{414})              &55.1\% (\emph{100\%})               \\
wrong direction from initial node & 125 (\emph{0})               &35 (\emph{0})               &28.0\%  (\emph{n/a})                 \\
right direction, before peak, 1-step Simply Dominant only      &236 (\emph{598})               &220 (\emph{598})               &93.2\% (\emph{100.0\%})                \\
right direction, before peak, 0-step Simply Dominant      &220 (\emph{506})               &213  (\emph{506})               &96.8\%   (\emph{100.0\%})             \\
at peak&95 (\emph{276})              &58  (\emph{276})               &61.0\%  (\emph{100.0\%})               \\
past peak&12 (\emph{0}) &  3 (\emph{0}) &25.0\% (\emph{n/a})          \\
\midrule
Overall &1,240 (1,932) &841 (1,932) &67.8\% (100.0\%) \\
\bottomrule
\end{tabular}
}
\caption{Summary statistics of characterization of nodes and frequency of action consistent with dominant strategy for the \OSPUM\ mechanism. Actual and predicted (in parenthesis) counts and rates are provided for each characterization. \label{tab:sumOSP}}
\end{table}

Since the theoretically ideal subject should never be off the equilibrium path---i.e., such subjects should never be moving away from their peak---the raw rates of node-level dominant strategy play suffer from a selection issue. Only subjects that deviate from the equilibrium path will be observed in off-equilibrium situations (i.e., past peak, or wrong direction from initial node). Thus the low rates of dominant-strategy play may not be due to the strategic complexity of the situation, just the selection of subjects. To remedy to this issue, Table \ref{fixedreg} provides a fixed effects regression of the rate of dominant-strategy actions on different types of node classifications. Importantly, the fixed effects specification calculates coefficients \emph{within subjects}, which should eliminate these selection issues provided they only exist at the subject level and are time-invariant.

The baseline (omitted) classification is a on-equilibrium, non-initial node, in the first half of the experiment, where the dominant-strategy action is continuation and is not part of a 0-step Simply Dominant strategic plan. Overall, subjects play the action associated with a dominant strategy in 90.6\% of these instances (97 of 107). An individual subject is 27 percentage points less likely to play such an action at their initial node ($p<0.05$). Subjects are also estimated to be 42 percentage points less likely to play the dominant-strategy action when it requires stopping ($p<0.01$).\footnote{This number should be thought of as a theoretical comparative static. In reality, all such cases would be part of a 0-step Simply Dominant strategic plan, so we must subtract 4.2 from this number to get a comparison relative to baseline ($-0.380$, $p<0.01$).} While the raw totals in Table \ref{tab:sumOSP} might suggest otherwise, rates of dominant strategy play do not differ greatly between off-equilibrium path nodes (where subjects must terminate) and being at one's peak (where subjects must also terminate). The former is estimated to be 4.2 percentage points lower than the latter, but the difference is not statistically significant ($p\approx0.594$). The value is far different from the 35-percentage point magnitude indicated in Table \ref{tab:sumOSP}; this difference suggests the regression is correcting some subject selection issues.
\begin{table}[]
\centering
\footnotesize
\begin{tabular}{lc} \hline
 & (1) \\
 & plays action \\
 & of dominant \\ 
 &  strategy path\\\hline
second half & 0.045 \\
 & (0.031) \\
initial node & -0.269** \\
 & (0.063) \\
off eq path & -0.045 \\
 & (0.064) \\
0-step simple & 0.042* \\
 & (0.015) \\
termination  & -0.422*** \\
 required & (0.062) \\ \hline
observations & 1,240 \\
number of subjects & 46 \\
r-squared & 0.155 \\ \hline
\end{tabular}
\caption{Fixed-effects regression of actions consistent with dominant strategy on node characterization\label{fixedreg}. Baseline observation is non-initial, one-step simple node on equilibrium path of play where continuation is associated with dominant strategy in the first half of the experiment. Valuation dummy variables are also included in regression. Standard errors are calculated from wild cluster bootstraps at the session level.}
\end{table}
%

Finally, our regression includes a term for nodes where the dominant-strategy action can be part of a 0-step Simply Dominant strategic plan. The estimated difference suggests this finer restriction increases the incidence of actions consistent with a peak-report by 4 percentage points ($p<0.10$). This result is consistent with Hypothesis \ref{Hyp:Py-T}. However, given the suggestive level of statistical significance and previously mentioned caveats concerning selection and specification, this result should be interpreted with caution.  

All in all, incentives seem to be affecting frequencies of play in the direction that theory predicts. Even the subtle differences between $0$-step Simple Dominance and $1$-step Simple Dominance are creating a measurable difference. However, the degree of success at the node level of \OSPUM\ is far from that necessary to surpass our benchmark \DRUM.

\subsection{Why \SRUM\ only mildly succeeds}
\begin{result}\label{res:SRUM} For the \SRUM :
\begin{enumerate}[label=\alph*.]
\item While rates of peak reporting do not differ between \SRUM1 and \DRUM\ subjects, reports for the former tend to be more erratic and deviate further from peak.
\item \SRUM1 subjects are no more likely to select reports consistent with a Perfect Equilibrium strategy than \DRUM\ subjects.
\item \SRUM1 subjects appear to harm efficiency more than corresponding subjects in \DRUM . However, the differential benefit of \SRUM2 over the \DRUM\ equivalent overcomes this deficiency.
\end{enumerate}
\end{result}

Result \ref{Res:individual} showed that the \SRUM1, \SRUM2, and \DRUM\ subjects all submit peak reports at the same rate---roughly 35\% of the time (see Table \ref{tab:indy1}). However, in terms of absolute deviations from peak, we see that subjects in \SRUM1 vary from the peak by 3.74 units on average compared to 2.79 in \DRUM\ ($p<0.05$), suggesting the dominant strategy property may lead to less noise in reports.\footnote{All significance values come from regressions similar in structure to those in Table \ref{reg:pair}, found in Appendix Tables \ref{reg:A1indy} and \ref{reg:A2indy}.} 

\begin{table}\footnotesize
\begin{centering}
\begin{tabular}{cccc}
\hline
valuation & peak & \begin{tabular}{c}SPE reports\\for \SRUM1 \end{tabular} & \begin{tabular}{c}Necessarily \\surplus-destroying\\ reports (\SDB )\end{tabular}  \\
\hline
1 & 3 & $\{0,\ldots ,10\}$ & $\{\emptyset \}$ \\
1 & 4 & $\{0,\ldots ,10\}$ & $\{\emptyset \}$\\
2 & 15 & $\{10,\ldots ,20\}$ & $\{\emptyset \}$ \\
2 & 16 & $\{10,\ldots ,20\}$ & $\{\emptyset \}$ \\
3 & 4 & $\{4\}$ & $\{5,\ldots ,20\}$ \\
3 & 16 & \{16\} & $\{0,\ldots ,15\}$ \\
4 & 3 & $\{0,\ldots ,7\}$ & $\{ 8,\ldots ,20 \}$ \\
4 & 13 & \{13\} & $\{ 0,\ldots ,13 \}$ \\
5 & 5 & $\{ 5 \}$ & $\{ 6,\ldots ,20 \}$ \\
5 & 17 & $\{15,\ldots ,20\}$  & $\{0,\ldots ,14\}$ \\
6 & 9 & $\{0,\ldots ,9\}$ & $\{10,\ldots ,20\}$ \\
6 & 11 & $\{11,\ldots ,20\}$ & $\{0,\ldots ,10\}$ \\
\hline
\end{tabular}
\par\end{centering}
\caption{Characterization of reports by peak and valuation pair.\label{tab:reportclass} Surplus destroying reports are those that when selected lead to inefficient outcomes regardless of the other player's strategy.}
\end{table}

A possible explanation is that \SRUM1 subjects are playing strategies that do not involve reporting one's peak, but are consistent a subgame perfect equilibrium. Table \ref{tab:reportclass} classifies the range of for each peak in all valuations. Indeed, in most situations, there are many possible strategies consistent with a subgame perfect equilibrium that do not involve reporting one's peak. However, \SRUM1 subjects play such strategies---non-peak reports that are consistent with an SPE---\emph{less} often than in \DRUM\ (25.8\% vs.\ 29.1\%, $p\approx0.365$), so this explanation cannot account for the greater deviation in reports.

To better understand how a non-peak report from \SRUM1 might hurt the outcomes of the \SRUM , we classify reports into those that \emph{necessarily} destroy surplus. 
Surplus destroying reports (\SDB ) are a key component to the overall performance of any of our mechanisms. Consider valuation 3 with peak pair (4, 16). If both players report their peaks, they will receive their most preferred allocation and utility totals 40. If instead, one subject reports 10, the allocation will be \{10, 10\}, regardless of the report of the other player. Total utility would be 28. Thus, a report of 10 is a \SDB ; it necessarily destroys 12 units of surplus.\footnote{The surplus destroyed by each players \SDB\ is generally not additive. For instance, if both subjects report 10, each necessarily destroys 12 of surplus, but only 12 total units of utility is lost in total.} Since the Uniform outcome is always efficient, by definition, any \SDB\ also prevents the Uniform outcome (though the converse is not true). 
Table \ref{tab:reportclass} characterizes the set of \SDB s for each peak and valuation. Note that in valuations 1 and 2, because the Uniform outcome is \{10,10\} no \SDB s exist: regardless of what one player reports, the other player can preserve the Uniform outcome (and full efficiency) by reporting 10. 

We will use the concept of \SDB\ to see whether the greater variability of reports by \SRUM1 differentially affects mechanism performance. Further we wish to show \SRUM2 reports can make up for this performance.
%
In general, \SDB\ are played at similar rates between \SRUM1 and \DRUM\ subjects (40.13\% vs.\ 39.42\%, respectively $p\approx 0.881$). It may be the case that reports in the \SRUM1 destroy a greater amount of surplus unilaterally, but we cannot show that at a conventional level of significance (2.855 vs. 2.366, $p\approx 0.152$). Nonetheless, we may investigate whether the performance of \SRUM2 (relative to \DRUM) is enough to make up this nominal deficit. 

The answer is decidedly yes. Suppose we imagine a player that strictly maximizes surplus given the report of their opponents. Relative to this ideal, we see that only an additional 0.553 units of utility per period is lost in the \SRUM2. In contrast, \DRUM\ subjects lose 1.67 units of utility (-1.120 difference, $p<0.01$). Combining these terms across regressions, we estimate a positive difference in differences ($1.120-0.490=0.631$, $p<0.1$), meaning the gains in efficiency (relative to \DRUM ) due to the \SRUM2 is enough to offset the losses in efficiency (relative to \DRUM ) of the \SRUM1.

\section{Discussion and concluding remarks}\label{Sec-Discussion}
This paper compared four mechanisms for a particular problem in mechanism design. These mechanisms are inspired by the major literatures in dominant strategy, robust, and obviously dominant mechanism design. We ranked the mechanisms both at the mechanism and node level using this literature. 
Variations at both the mechanism and node level allowed us to test the various proposed comparative statics in this literature. 

At the mechanism level, a dominant strategy mechanism that provides feedback during a reporting period is the best performer in terms of the frequency of desirable outcomes obtained in our experiment. Our data suggests that providing feedback about tentative partner reports made it easier for subjects to maximize payoffs and encouraged them to employ dominant strategies. It may be possible to obtain similar improvements by providing feedback during the reporting period in other direct revelation mechanisms. As computational power increases, this becomes technically feasible in field applications. Perhaps more important than the ways in which our feedback-augmented mechanism succeeded, are the ways in which other mechanisms did not succeed or failed. A main concern is the exponential accumulation of mistakes at the node level in mechanism that require long equilibrium paths. These issues may present serious risks for implementation. 

Our results identifying the driving forces behind the differential performance of our mechanisms allow us to make an informed guess about the validity of these comparisons for situations that involve more players. Since coordination in \DRUM\ becomes more difficult as the number of players increases, one can expect that its performance would not improve. Since the slight increase of performance of \SRUM\ relied on the second mover, one can expect that this effects gets diluted as the number of players increases. By contrast, the advantages of \DFUM\ have the potential to survive with more players. No matter the number of players, subjects will have a concrete informative conjecture to which they can best respond. Variation in tentative reports may continue to flag dominant strategies as superior. Thus, one can expect higher best response rates and more frequent truthful reports in \DRUM. For Uniform rationing, an increase in truthful play among best responders increases performance because efficiency of equilibria requires only that the agent with the most extreme peak reports truthfully \citep{BOCHET-Tumme-JET-2020}.  Since the number of moves necessary to obtain the desired outcomes in \OSPUM\ does not decrease when there are more agents, we cannot expect that the performance of this mechanism would improve with $n>2$.

\bibliography{ref-UR}
\appendix
\renewcommand{\thetable}{A.\arabic{table}}
\renewcommand\thefigure{A.\arabic{figure}}
\renewcommand\theequation{A.\arabic{equation}}
\setcounter{equation}{0}
\setcounter{table}{0}
\setcounter{figure}{0}

\break
\section{Additional Tables and Figures}

\begin{table}[h]
\centering
\begin{threeparttable}\footnotesize
\begin{tabular}{lcccc}
&(1)&(2)&(3)&(4)\\
&\begin{tabular}{c}Uniform\\ outcome\end{tabular}&\begin{tabular}{c}efficient\\ outcome\end{tabular}&\begin{tabular}{c}within-1\\ of efficient\\ outcome\end{tabular}&\begin{tabular}{c}share of\\ efficiency\end{tabular}\\\hline
second half&0.052*&0.030&0.048&0.011\\
&(0.028)&(0.022)&(0.030)&(0.010)\\
\DFUM\ &0.249***&0.235***&0.190***&0.038***\\
&(0.051)&(0.058)&(0.037)&(0.011)\\
\OSPUM\ &-0.002&0.103***&0.078***&0.015**\\
&(0.043)&(0.035)&(0.029)&(0.007)\\
\SRUM\ &0.052*&0.107***&0.075**&0.009\\
&(0.028)&(0.030)&(0.033)&(0.007)\\
\OSPUM\ $\times $&-0.070&-0.179**&-0.194***&-0.053***\\
valuations 3--6 &(0.090)&(0.080)&(0.070)&(0.018)\\
\hline
observation level\tnote{a}&decision-pair&decision-pair&decision-pair&decision-pair\\
observations&960&960&960&960\\
log likelihood&-586.936&-441.877&-347.989&1007.226\\\hline
\end{tabular}
\begin{tablenotes}
\item \emph{Notes:} \DRUM\ is baseline model. Valuation dummy variables are also included in all regressions. All regression models use cluster-robust standard errors at the session level.
\end{tablenotes}
\end{threeparttable}
\caption{Regression analysis of pair-level outcomes by mechanism with added interaction term.\label{reg:pair-noeq}}
\end{table}

\begin{table}\footnotesize
\centering
\begin{tabular}{cccccc}
\toprule
&share of& \multicolumn{4}{c}{\multirow{2}{*}{rate of Uniform outcome observed}}\\
valuation& profiles that & & & & \\
&produce $U$& \DRUM\ & \SRUM\ & \OSPUM\ & \DFUM\  \\
\midrule
1 & 0.546& 0.788 & 0.684 & 0.696  & 0.875 \\
2 & 0.546& 0.865 &0.632 & 0.739 & 0.958 \\
3 & 0.020& 0.269 & 0.421 & 0.196 & 0.583  \\
4 & 0.034& 0.423 & 0.395 & 0.239 & 0.625 \\
5 & 0.025& 0.212 & 0.474 & 0.261  & 0.583 \\
6 & 0.043& 0.365 & 0.632 & 0.500  & 0.792  \\
\midrule
overall & 0.203 & 0.487 & 0.539 & 0.438 & 0.736 \\
\midrule
&share of& \multicolumn{4}{c}{\multirow{2}{*}{rate of Nash profile observed }}\\
valuation&profiles that& \\
&are Nash& \DRUM\ & \SRUM\ & \OSPUM\ & \DFUM\  \\
\midrule
1&0.274&0.673&0.553&0.913&0.875\\
2&0.274&0.808&0.579&0.913&0.917\\
3&0.034&0.288&0.658&0.326&0.667\\
4&0.025&0.442&0.605&0.413&0.667\\
5&0.025&0.231&0.711&0.348&0.625\\
6&0.045&0.404&0.789&0.652&0.917\\
\midrule
overall & 0.113&0.474&0.649&0.594&0.778\\
\midrule
&share of& \multicolumn{4}{c}{\multirow{2}{*}{rate of truthful profile observed }}\\
valuation&profiles that are&\\
& truthful reports& \DRUM\ & \SRUM\ & \OSPUM\ & \DFUM\  \\
\midrule
1&0.002&0.077&0.079&0.261&0.042\\
2&0.002&0.096&0.053&0.391&0.083\\
3&0.002&0.231&0.237&0.000&0.542\\
4&0.002&0.135&0.211&0.250&0.167\\
5&0.002&0.096&0.105&0.239&0.167\\
6&0.002&0.212&0.421&0.043&0.750\\
\midrule
overall & 0.002&0.141&0.184&0.197&0.292\\
\bottomrule
\end{tabular}
\caption{Share of profile space that produces Uniform outcome (top), that is Nash (middle), and the truthful profile (bottom). Actual rates of observed profile for each mechanism and valuation are provided. \label{tab:Aprofiles}}
\end{table}
\begin{table}[t]
\centering
\begin{threeparttable}\footnotesize
\centering
\begin{tabular}{@{}p{0.9\textwidth}@{}}
\centering
\begin{tabular}{lcccc} \hline
 & (1) & (2) & (3) & (4) \\
 & & plays & absolute & non-peak \\
 & reports & best & deviation &  SPE\\
 & peak & response & from peak & report\tnote{a} \\ \hline
second half & 0.056** & 0.021 & -0.338* & -0.011 \\
 & (0.020) & (0.024) & (0.171) & (0.022) \\
\DFUM\ & 0.151** & 0.162*** & -0.729** & 0.019 \\
 & (0.054) & (0.024) & (0.332) & (0.019) \\
\SRUM1  & -0.006 & -0.108*** & 0.939** & -0.034 \\
& (0.043) & (0.028) & (0.424) & (0.037) \\
\SRUM2 & 0.016 & 0.142*** & 0.404 & -0.013 \\
& (0.046) & (0.017) & (0.394) & (0.019) \\
\OSPUM\  & 0.104** & -0.094*** & -0.718* & -0.034 \\
path level\tnote{b}  & (0.048) & (0.026) & (0.340) & (0.027) \\
\hline
observation level& strategy& strategy& strategy &strategy\\
observations & 1,920 & 1,920 & 1,920 & 1,920 \\
 r-squared & 0.046 & 0.071 & 0.103 & 0.200 \\ \hline
\end{tabular}
\end{tabular}
\begin{tablenotes}
\item \emph{Notes:} \DRUM\ is baseline model. Valuation dummy variables are also included in all regressions. All regression models use cluster-robust standard errors at the session level. Regressions (1) and (2) are also found in Table \ref{reg:indy}.
\item[a] Regardless of mechanism, report falls within the set of subgame perfect equilibrium reports for the first mover in \SRUM , excluding peak reports. See Table \ref{tab:reportclass} for a full classification of such reports.
\item[b] Report is assumed to be a peak report unless subject chose to terminate in a way inconsistent with such strategy. In those cases, the report is considered to be value in which the subject chose to terminate.
\end{tablenotes}
\end{threeparttable}
\caption{Regression analysis of individual-level decisions by mechanism and role.\label{reg:A1indy}}
\end{table}
\begin{table}[t]
\centering
\begin{threeparttable}\footnotesize
\centering
\begin{tabular}{@{}p{0.9\textwidth}@{}}
\centering
\begin{tabular}{lcccc} \hline
 & (1) & (2) & (3) & (4) \\
 & & surplus  &  & difference  \\
 & plays &  necessarily & payoff  & from \SDB\\
 & \SDB\tnote{a} & lost from  & received  & to payoff\tnote{b} \\ 
 &  &report  &  &  \\ \hline
second half & -0.062** & -0.438** & 0.209 & 0.019 \\
 & (0.023) & (0.189) & (0.143) & (0.149) \\
\DFUM\ & -0.202*** & -1.272*** & 1.092*** & -0.913*** \\
 & (0.060) & (0.398) & (0.295) & (0.219) \\
\SRUM1 & 0.007 & 0.490 & 0.256 & -1.120*** \\
& (0.046) & (0.322) & (0.190) & (0.136) \\
\SRUM2 & -0.026 & -0.089 & 0.374* & -0.541** \\
& (0.035) & (0.324) & (0.202) & (0.220) \\
\OSPUM\ & 0.003 & 0.520 & -0.606 & 0.691** \\
path level\tnote{c}  & (0.057) & (0.490) & (0.379) & (0.311) \\
\hline
observation level& strategy& strategy& strategy &strategy\\
observations & 1,280 & 1,280 & 1,280 & 1,280 \\
r-squared & 0.039 & 0.134 & 0.271 & 0.097 \\ \hline
\end{tabular}
\end{tabular}
\begin{tablenotes}
\item \emph{Notes:} Necessarily surplus destroying-reports (\SDB ) are only possible in valuations 3--6, hence only those valuations are used. This reduces total observations from 1,920 to 1,280. \DRUM\ is baseline model. Valuation dummy variables are also included in all regressions. All regression models use cluster-robust standard errors at the session level.
\item[a] Plays a report that \emph{necessarily} leads to inefficient outcomes regardless of the other player's strategy. See full characterization in Table \ref{tab:reportclass}.
\item[b] This is the difference between actual loss of surplus from full efficiency and the surplus necessarily destroyed by the report of one's opponent. Note that the value must always be non-negative.
\item[c] Observed strategy paths consistent with peak reports.
\end{tablenotes}
\end{threeparttable}
\caption{Regression analysis of individual-level decisions by mechanism and role.\label{reg:A2indy}}
\end{table}

\end{document}